\documentclass{article}

\usepackage{booktabs} 

\usepackage[normalem]{ulem}

\usepackage{epsfig}
\usepackage{cite}
\usepackage{amsmath}
\usepackage[footnotesize]{caption}
\usepackage{slashed}
\usepackage{xcolor}
\usepackage[utf8]{inputenc}
\usepackage{tikz}
\usepackage{tikz-feynman}
\usetikzlibrary{shapes.geometric}
\usetikzlibrary{arrows.meta,arrows}
\usetikzlibrary{quotes,angles}
\usepackage{subcaption}
\usepackage{amssymb}
\usepackage{multicol}
\usepackage{multirow}
\usepackage{array}

\usepackage{cancel}

\numberwithin{equation}{section}

\newcommand{\be}{\begin{equation}}
\newcommand{\ee}{\end{equation}}
\newcommand{\beq}{\begin{eqnarray}}
\newcommand{\eeq}{\end{eqnarray}}

\usepackage{comment}

\usepackage[english]{babel}
\usepackage{indentfirst}

\usetikzlibrary{external}
\begin{document}

\title{Precision Muon-Related Observables as a Tool\\ to Constrain New Physics Models}

\date{\today}
\author{
Gabriel Lourenço$^{1\,}$\footnote{E-mail:
\texttt{gabaslourenco@gmail.com}} ,
Andre Milagre$^{2\,}$\footnote{E-mail:
\texttt{andre.milagre@gmail.com}} ,
Rui Santos$^{1,3\,}$\footnote{E-mail:
  \texttt{rasantos@fc.ul.pt}} ,
João P. Silva$^{2\,}$\footnote{E-mail:
\texttt{jpsilva@cftp.ist.utl.pt}} ,
\\[5mm]
{\small\it $^1$Centro de F\'{\i}sica Te\'{o}rica e Computacional,
    Faculdade de Ci\^{e}ncias,} \\
{\small \it    Universidade de Lisboa, Campo Grande, Edif\'{\i}cio C8
  1749-016 Lisboa, Portugal} \\[3mm]
{\small\it $^2$ CFTP, Departamento de F\'{\i}sica,} \\
{\small\it Instituto Superior T\'{e}cnico, Universidade de Lisboa,} \\
{\small\it Avenida Rovisco Pais 1, 1049-001 Lisboa, Portugal} \\[3mm]
{\small\it
$^3$ISEL -
 Instituto Superior de Engenharia de Lisboa,} \\
{\small \it   Instituto Polit\'ecnico de Lisboa
 1959-007 Lisboa, Portugal} \\[3mm]
}

\maketitle

\begin{abstract}
\noindent
We propose a set of precision muon-related observables that serve as a tool to constrain new physics models. Using LEP's precision measurements 
on the $Z$-boson pole, we derive bounds on the new physics quantum contributions to the decay $Z \to \mu^+ \mu^-$. We show that the new precision observables 
have a real impact on two specific models that solve the $g-2$ anomaly and provide a sound dark matter candidate.

\end{abstract}

\thispagestyle{empty}
\vfill
\newpage
\setcounter{page}{1}

\section{Introduction}

Until 2022 there was an observed anomaly in the semileptonic $B$ meson decay rates that suggested a violation of lepton flavor universality. 
However, the recent reinterpretation performed by the LHCb collaborations washed out the discrepancy between the SM prediction in the $b\to s\mu^+\mu^-$ transition~\cite{LHCb:2022vje, LHCb:2022qnv}
and the experimentally measured value. The anomaly was present for a long time in the measurements of the ratios of the exclusive branching fractions, $R(K^{(*)}) = {\cal B}(B\to K^{(*)}\mu^+\mu^-)/{\cal B}(B \to K^{(*)}e^+ e^-)$, 
obtained by the LHCb Collaboration~\cite{LHCb:2021trn,LHCb:2019hip,LHCb:2017avl} and by the Belle Collaboration~\cite{Belle:2019oag,BELLE:2019xld} although with larger errors. In an attempt to solve this problem, several new models 
were proposed among which some focused on adding a number of new fields that would solve the problems via one-loop contributions~\cite{Bhattacharya:2015xha, Kawamura:2017ecz, Cline:2017qqu, Cerdeno:2019vpd, Barman:2018jhz, 
Darme:2020hpo, Arcadi:2021cwg, Becker:2021sfd, PhysRevD.102.075009, Capucha:2022kwo} from particles living in a new Dark Sector. Some of these models~\cite{Cerdeno:2019vpd, PhysRevD.102.075009, Capucha:2022kwo} also solved the muon 
$g-2$ anomaly~\cite{Muong-2:2023cdq, ParticleDataGroup:2018ovx,Gorringe:2015cma, Aoyama:2020ynm, Muong-2:2006rrc}  while providing a Dark Matter (DM) candidate compatible with all present constraints. 
It is important to note that although the anomaly in the semileptonic  $B$ meson decay rates vanished, the latter models are still in agreement with all experimental data and are able to explain the $g-2$ anomaly and still
provide a scalar dark matter candidate.

Triggered by exploring in more detail the constraints on this class of models we thought it was timely to use the precision measurement of muon-related observables to constrain any new physics contributions coming from the proposed extensions
of the  Standard Model (SM). 
The idea is inspired by the use of precision observables related to $Z \to b \bar b$~\cite{Field:1997gz, PhysRevD.62.015011, Fontes:2019fbz, Jurciukonis:2021wny} to constrain models with an extended scalar sector of the SM. In fact, the precision
measurements performed at LEP on the $Z$-boson pole can be used for other $Z$ decay channels such as $Z \to \mu^+ \mu^-$ and $Z \to \tau^+ \tau^-$. If a new Dark Sector contributes to a new observable like for instance the muon $g-2$, it will certainly appear in other observables related to the muon.
The need to have large couplings to muons will then manifest itself in the $Z$ decays to muon allowing to restrict the parameter space of the model.
We will further show that the results for the tau lepton are obtained directly from the muon and that because the experimental measurements related to the tau have larger error bars, the results are less restrictive in constraining the parameter space of a new model.

The paper is organized as follows. In Sec.~\ref{sec:obs} we present and discuss the $Z \to \mu^+ \mu^-$ observables to be used and in Sec.~\ref{radcor} we present the calculation of the radiative corrections. In Sec.~\ref{sec:results} we 
present the results and apply the constraints to two models of the new physics.
Our conclusions are given in Sec.~\ref{sec:conclusions}.

\section{Observables}
\label{sec:obs}

In the  SM, the interaction Lagrangian between the $Z$ and the muons can be written as
\begin{equation}
	\mathcal{L}_{Z\mu^+\mu^-} = -\frac{g}{c_W} Z^\lambda \bar{\mu}\, \gamma_\lambda \left( g_{\mu L} P_L + g_{\mu R} P_R \right) \mu,
\end{equation}
where $P_{L}$ and $P_R$ are the left and right-handed chirality projectors. At tree-level, the $Z\mu^+\mu^-$ couplings in the SM are
\begin{align}
	&& g_{\mu L} &= -\frac{1}{2} + s_W^2 ,& g_{\mu R} &= s_W^2 , && 
\end{align}

\noindent where $s_W$ and $c_W$ are the sine and cosine of the Weinberg angle, respectively. These quantities were measured at LEP~\cite{ALEPH:2005ab}  
and the most precise values for the corresponding SM couplings read~\cite{ALEPH:2005ab} 
\begin{align}
	&& g_{\mu L}^{\textsc{SM}} &= -0.26929 ,& g_{\mu R}^{\textsc{SM}} &= 0.23208. &&
	\label{eq:SMg}
\end{align}
As previously discussed, quantum loop corrections make the couplings deviate from their tree-level values. 
These corrections are partly generated by SM particles and partly from the new physics contributions.

The muon couplings can be derived from two $Z$-pole observables related to the $e^+ e^- \to \mu^+ \mu^-$ process: the asymmetry parameter ($\mathcal{A}_\mu^0$) and the muon hadronic decay rate ($R_\mu$). 
Henceforth, we assume all observables to be corrected for initial-state radiation but make the distinction between observables that include effects of final-state radiation by placing a "0" superscript on the ones that do not. 

The asymmetry parameter can be calculated from the muon forward-backward asymmetry $A_\textup{FB}^0$. This is a measure of the angular asymmetry in the distribution between final state particles produced in the \textit{forward} direction (defined by the angle they make with the initial electron beam, \textit{i.e.} $\theta < \frac{\pi}{2}$) and the ones that are produced \textit{backward} ($\theta > \frac{\pi}{2}$). Its formal definition reads
\begin{equation}
	A_\textup{FB}^0 = \frac{\sigma_F^0-\sigma_B^0}{\sigma_F^0-\sigma_B^0} = \frac{3}{4} \mathcal{A}_e^0 \mathcal{A}_\mu^0,
\end{equation}
 with,
\begin{align}
	\sigma_F^0 &= 2\pi \int_0^{1} d\cos \theta \, \frac{d\sigma^0}{d \, \cos \theta} , & \sigma_B^0 &= 2\pi \int_{-1}^{0} d\cos \theta \, \frac{d\sigma^0}{d \, \cos \theta}.
\end{align}
which are the cross sections of the muons detected in the forward and backward directions, respectively. Since the center of mass energy range is near the $Z$ resonance ($s \sim m_Z^2$), $\mathcal{A}_\mu^0$ can be written in terms of the tree-level left and right $Z\mu^+ \mu^- $ couplings as
\begin{equation}
	\mathcal{A}_\mu^0 \equiv  \frac{\sqrt{1-4 \mu_\mu} (g_{\mu L}^{2}-g_{\mu R}^{2})}{ (1-\mu_f)(g_{\mu L}^{2} + g_{\mu R}^{2}) + 6 \mu_\mu\, g_{\mu L}\, g_{\mu R}  }\ \ \ ,\ \ \ \mu_\mu \equiv \frac{m_\mu^2}{m_Z^2}.
	\label{eq:Amu}
\end{equation}
where $m_\mu$ is the mass of the muon and $m_Z$ that of the $Z$ boson. 

The muon hadronic decay rate is defined as the ratio between the partial decay width of the $Z$ to hadrons and to muons,
\begin{align}
	R_\mu^0 &\equiv \frac{\Gamma^0(Z \to had)}{\Gamma^0(Z \to \mu^+ \mu^-)} , & \Gamma^0(Z \to had)&= \sum_{q=u,d,c,s,b} \Gamma^0(Z \to q \bar{q}),
\end{align}
which depends on the couplings between the quarks and the $Z$. As will be explained in Section \ref{radcor}, although cross-sections are defined experimentally without final-state corrections, partial widths are not~\cite{ALEPH:2005ab}. 
To avoid performing QCD-related computations for the final state radiation of quarks, instead of working with $R_\mu^0$ directly we define the quantity~\footnote{This observable was used to constrain the parameter space of a Leptophilic two-Higgs doublet model~\cite{Chun:2016hzs}.} $R_{e\mu}=R_e/R_\mu$ which, at tree level, and without final state radiation factors, can be written as
\begin{equation}
	R_{e\mu}^0 \equiv \frac{R_e^0}{R_\mu^0} = \frac{\Gamma^0(Z \to \mu^+ \mu^-)}{\Gamma^0(Z \to e^+ e^-)} = \frac{  g_{\mu L}^{2} + g_{\mu R}^{2} -\mu_\mu (g_{\mu L}^{2} + g_{\mu R}^{2} - 6\,g_{\mu L}\,g_{\mu R}) }{  g_{e L}^{2} + g_{e R}^{2} }.
	\label{eq:Rem}
\end{equation}

There is good agreement between the predicted and experimental results of both $\mathcal{A}_\mu^0$ and $R_{e\mu}$ since the central SM value is in the uncertainty interval of the corresponding experimental result. Table \ref{table:compare_SM} presents the corresponding SM prediction and fitted values for $R_e$, $R_\mu$, $\mathcal{A}^0_\mu$, and $R_{e\mu}$. The uncertainty in $R_{e\mu}$ was calculated from the quadratic sum of the uncertainties in $R_e$ and $R_\mu$, since these observables are poorly correlated \cite{ALEPH:2005ab}.
\begin{table}[ht]

\centering
\begin{tabular}{c c c }
\hline
\hline
Observable               & Exp                  & SM                  \\ \hline
$R_e$                    & $20.804 \pm 0.050$   & $20.736 \pm 0.010$  \\ 
$R_\mu$                  & $20.784 \pm 0.034$   & $20.736 \pm 0.010$  \\
$\mathcal{A}^0_\mu$      & $0.142 \pm 0.015$    & $0.1468 \pm 0.0003$ \\
$R_{e\mu}$               & $1.0009 \pm 0.0029$  & $1.0000 \pm 0.0007$ \\ \hline
\end{tabular}
\caption{Experimental (Exp) value and Standard Model (SM) predictions for $R_e$, $R_\mu$, $\mathcal{A}_\mu^0$ and $R_{e\mu}$ \cite{ParticleDataGroup:2022pth}. The SM predictions include the one-loop contributions, as well as the dominant two-loop contributions. $R_{e\mu}$ is obtained from the hadronic decay rates of the electron and the muon in the first two lines.}
\label{table:compare_SM}
\end{table}

The SM predictions include one-loop contributions and the dominant two-loop contributions as described in the review Electroweak Model and Constraints on New Physics~ \cite{ParticleDataGroup:2022pth}.

\section{Radiative corrections}
\label{radcor}

\begin{figure}[h!]
\begin{tikzpicture}[scale=0.73]
    \begin{feynman}
      \vertex (a) at (-2.07, 1.1);
      \vertex (am) at (-1.04,0.37);
      \vertex (b) at (-2.07, -1.1);
      \vertex (v1) at (-0.6,0);
      \vertex (v2) at (1.5,0);
      \vertex (c) at (2.97,1.1);
      \vertex (d) at (2.97,-1.1);
      \vertex (g) at (0.3,0.9);
      \diagram* {
        (a) --[fermion, edge label=$e^-$] (am),
        (am)--[fermion] (v1),
        (am)--[photon,edge label=$\gamma$] (g),
        (b) --[anti fermion, edge label'=$e^+$] (v1),
        (v1) --[photon, edge label'={$\gamma,Z$}] (v2),
        (v2)--[fermion,edge label=$\mu^-$] (c),
        (v2)--[anti fermion,edge label'=$\mu^+$] (d),
        };
    \end{feynman}
\end{tikzpicture}
\hfill
\hfill
\begin{tikzpicture}[scale=0.73]
    \begin{feynman}
      \vertex (a) at (-2.07, 1.1);
      \vertex (b) at (-2.07, -1.1);
      \vertex (bm) at (-1.04,-0.37);
      \vertex (v1) at (-0.6,0);
      \vertex (v2) at (1.5,0);
      \vertex (c) at (2.97,1.1);
      \vertex (d) at (2.97,-1.1);
      \vertex (g) at (0.3,-0.9);
      \diagram* {
        (a) --[fermion, edge label=$e^-$] (v1),
        (b) --[anti fermion, edge label'=$e^+$] (bm),
        (bm)--[anti fermion] (v1),
        (bm)--[photon,edge label'=$\gamma$] (g),
        (v1) --[photon, edge label={$\gamma,Z$}] (v2),
        (v2)--[fermion,edge label=$\mu^-$] (c),
        (v2)--[anti fermion,edge label'=$\mu^+$] (d),
        };
    \end{feynman}
\end{tikzpicture}
\hfill
\hfill
\begin{tikzpicture}[scale=0.73]
    \begin{feynman}
      \vertex (a) at (-2.07, 1.1);
      \vertex (b) at (-2.07, -1.1);
      \vertex (v1) at (-0.6,0);
      \vertex (v2) at (1.5,0);
      \vertex (c) at (2.97,1.1);
      \vertex (d) at (2.97,-1.1);
      \vertex (dm) at (1.99,-0.37);
      \vertex (g) at (2.97,-0.37);
      \diagram* {
        (a) --[fermion, edge label=$e^-$] (v1),
        (b) --[anti fermion, edge label'=$e^+$] (v1),
        (v1) --[photon, edge label={$\gamma,Z$}] (v2),
        (v2)--[fermion,edge label=$\mu^-$] (c),
        (v2)--[anti fermion] (dm),
        (dm)--[anti fermion, edge label'=$\mu^+$] (d),
        (dm)--[photon,edge label=$\gamma$] (g),
        };
    \end{feynman}
\end{tikzpicture}
\hfill
\hfill
\begin{tikzpicture}[scale=0.73]
    \begin{feynman}
      \vertex (a) at (-2.07, 1.1);
      \vertex (b) at (-2.07, -1.1);
      \vertex (v1) at (-0.6,0);
      \vertex (v2) at (1.5,0);
      \vertex (c) at (2.97,1.1);
      \vertex (cm) at (1.99,0.37);
      \vertex (g) at (2.97,0.37);
      \vertex (d) at (2.97,-1.1);
      \diagram* {
        (a) --[fermion, edge label=$e^-$] (v1),
        (b) --[anti fermion, edge label'=$e^+$] (v1),
        (v1) --[photon, edge label={$\gamma,Z$}] (v2),
        (v2)--[fermion] (cm),
        (cm)--[photon,edge label'=$\gamma$] (g),
        (cm)--[fermion,edge label=$\mu^-$] (c),
        (v2)--[anti fermion,edge label'=$\mu^+$] (d),
        };
    \end{feynman}
\end{tikzpicture}

\hfill

\begin{tikzpicture}
    \begin{feynman}[scale=0.73]
      \vertex (a) at (-2.07, 1.1);
      \vertex (b) at (-2.07, -1.1);
      \vertex (v1) at (-0.6,0);
      \vertex (v2) at (1.5,0);
      \vertex (c) at (2.97,1.1);
      \vertex (cm) at (2.235,0.55);
      \vertex (dm) at (2.235,-0.55);
      \vertex (d) at (2.97,-1.1);
      \diagram* {
        (a) --[fermion, edge label=$e^-$] (v1),
        (b) --[anti fermion, edge label'=$e^+$] (v1),
        (v1) --[photon, edge label={$\gamma,Z$}] (v2),
        (v2)--[fermion] (cm),
        (cm)--[fermion,edge label=$\mu^-$] (c),
        (v2)--[anti fermion] (dm),
        (dm)--[anti fermion,edge label'=$\mu^+$] (d),
        (cm)--[photon, edge label=$\gamma$] (dm)
        };
    \end{feynman}
\end{tikzpicture}
\hfill
\begin{tikzpicture}
    \begin{feynman}[scale=0.73]
      \vertex (a) at (-2.07, 1.1);
      \vertex (am) at (-1.235,0.55);
      \vertex (bm) at (-1.235,-0.55);
      \vertex (b) at (-2.07, -1.1);
      \vertex (v1) at (-0.6,0);
      \vertex (v2) at (1.5,0);
      \vertex (c) at (2.97,1.1);
      \vertex (d) at (2.97,-1.1);
      \diagram* {
        (a) --[fermion, edge label=$e^-$] (am),
        (am)--[fermion] (v1),
        (b) --[anti fermion, edge label'=$e^+$] (bm),
        (bm)--[anti fermion] (v1),
        (am)--[photon, edge label'=$\gamma$] (bm),
        (v1) --[photon, edge label={$\gamma,Z$}] (v2),
        (v2)--[fermion,edge label=$\mu^-$] (c),
        (v2)--[anti fermion,edge label'=$\mu^+$] (d)
        };
    \end{feynman}
\end{tikzpicture}
\hfill
\begin{tikzpicture}[scale=0.9]
    \begin{feynman}
      \vertex (a) at (0,0);
      \vertex (v1) at (1.3,0);
      \vertex (v2) at (2.6,0);
      \vertex (b) at (4,0);
      \vertex (v3) at (1.3,1.5  );
      \vertex (v4) at (2.6,1.5);
      \vertex (c) at (0,1.5);
      \vertex (d) at (4,1.5);
      
      \diagram* {
        (a) --[anti fermion,edge label'=$e^+$] (v1),
        (v1)--[boson, edge label'={$\gamma,\gamma,Z$}] (v2),
        (v2)--[anti fermion, edge label'=$\mu^+$] (b),
        (v1) --[anti fermion, edge label=$e$] (v3),
        (v2)--[fermion ,edge label'={{$\mu$}}] (v4),
        (v3)--[boson,edge label={$\gamma,Z,\gamma$}] (v4),
        (c) --[fermion, edge label=$e^-$] (v3),
        (v4) --[fermion, edge label=$\mu^-$] (d),        
        };
    \end{feynman}
\end{tikzpicture}
\hfill
\begin{tikzpicture}[scale=0.9]
    \begin{feynman}
      \vertex (a) at (0,0);
      \vertex (v1) at (1.3,0);
      \vertex (v2) at (2.6,0);
      \vertex (b) at (4,0);
      \vertex (v3) at (1.3,1.5);
      \vertex (v4) at (2.6,1.5);
      \vertex (c) at (0,1.5);
      \vertex (d) at (4,1.5);
      
      \diagram* {
        (a) --[anti fermion,edge label'=$e^+$] (v1),
        (v1)--[boson] (v4),
        (v2)--[anti fermion, edge label'=$\mu^+$] (b),
        (v1) --[anti fermion, edge label=$e$] (v3),
        (v2)--[fermion ,edge label'={{$\mu$}}] (v4),
        (v3)--[boson] (v2),
        (c) --[fermion, edge label=$e^-$] (v3),
        (v4) --[fermion, edge label=$\mu^-$] (d),        
        };
    \end{feynman}
    \node at (2,1.7) {$\gamma,\gamma,Z$};
    \node at (2,-0.2) {$\gamma,Z,\gamma$};
\end{tikzpicture}
\caption{QED-type diagrams contributing to the $e^+e^-\to \mu^+\mu^-$ process.}
\label{fig:QED_ee_mm}
\end{figure}
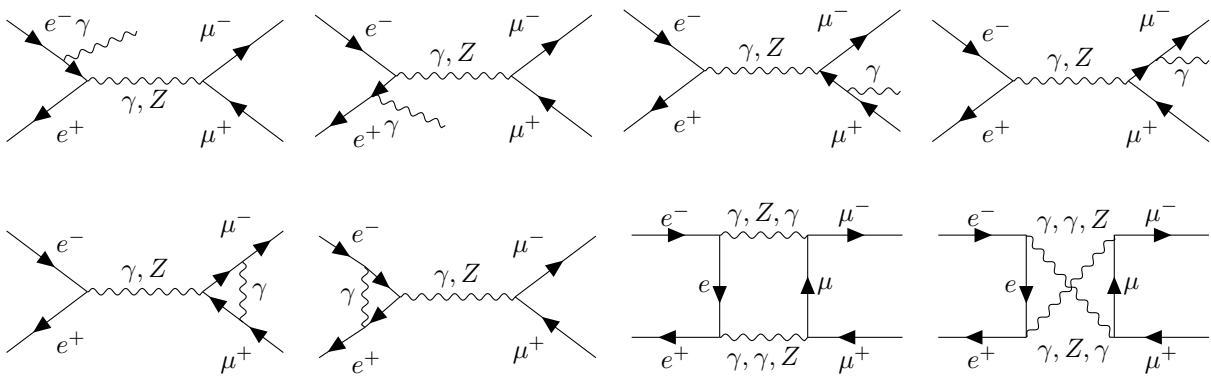

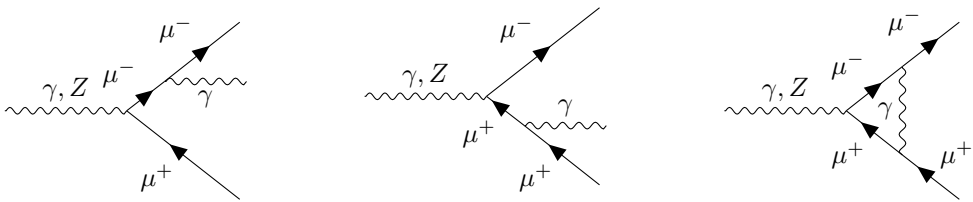
\begin{figure}
\hfill
\begin{tikzpicture}[scale=0.90]
    \begin{feynman}
      \vertex (a) at (-0.3,0);
      \vertex (v1) at (1.5,0);
      \vertex (b) at (3.17,1.3);
      \vertex (bm) at (2.06,0.43);
      \vertex (g) at (3.27,0.43);
      \vertex (c) at (3.17,-1.3);
      \diagram* {
        (a) --[photon, edge label={$\gamma,Z$}] (v1),
        (v1)--[fermion,edge label=$\mu^-$] (bm),
        (bm)--[fermion,edge label=$\mu^-$] (b),
        (bm)--[photon, edge label'=$\gamma$] (g),
        (v1)--[anti fermion,edge label'=$\mu^+$] (c)
        };
    \end{feynman}
\end{tikzpicture}
\hfill
\begin{tikzpicture}[scale=0.90]
    \begin{feynman}
      \vertex (a) at (-0.3,0);
      \vertex (v1) at (1.5,0);
      \vertex (b) at (3.17,1.3);
      \vertex (c) at (3.17,-1.3);
      \vertex (cm) at (2.06,-0.43);
      \vertex (g) at (3.27,-0.43);
      \diagram* {
        (a) --[photon, edge label={$\gamma,Z$}] (v1),
        (v1)--[fermion,edge label=$\mu^-$] (b),
        (v1)--[anti fermion,edge label'=$\mu^+$] (cm),
        (cm)--[anti fermion,edge label'=$\mu^+$] (c),
        (cm)--[photon, edge label=$\gamma$] (g),
        };
    \end{feynman}
\end{tikzpicture}
\hfill
\begin{tikzpicture}[scale=0.90]
    \begin{feynman}
      \vertex (a) at (-0.3,0);
      \vertex (v1) at (1.5,0);
      \vertex (b) at (3.17,1.3);
      \vertex (bm) at (2.33,0.65);
      \vertex (c) at (3.17,-1.3);
      \vertex (cm) at (2.33,-0.65);
      \diagram* {
        (a) --[photon, edge label={$\gamma,Z$}] (v1),
        (v1)--[fermion,edge label=$\mu^-$] (bm),
        (bm)--[fermion,edge label=$\mu^-$] (b),
        (v1)--[anti fermion,edge label'=$\mu^+$] (cm),
        (cm)--[anti fermion, edge label=$\mu^+$] (c),
        (bm)--[photon,edge label'=$\gamma$] (cm)
        };
    \end{feynman}
\end{tikzpicture}
\hfill
\hfill

\caption{QED-type diagrams contributing to the $Z\to \mu^+\mu^-$ process.}
\label{fig:QED_Z_mm}
\end{figure}

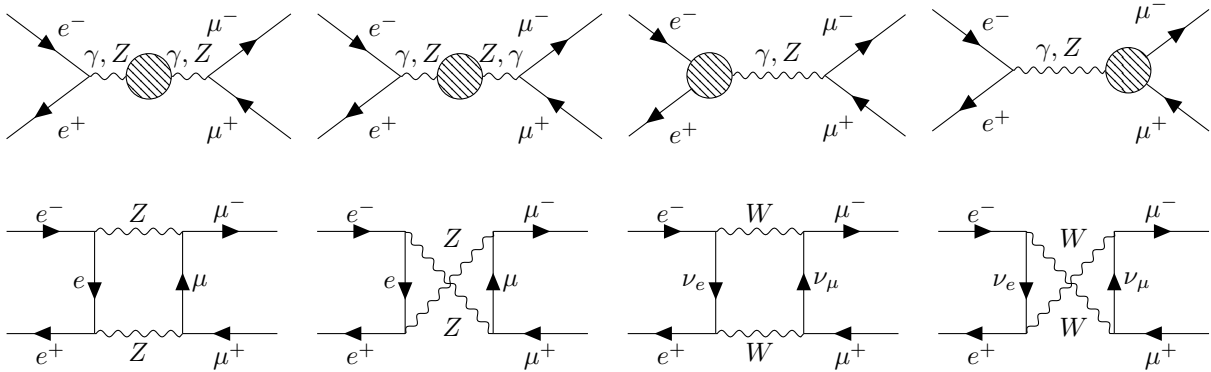
\begin{figure}
\begin{tikzpicture}[scale=0.75]
    \begin{feynman}
      \vertex (a) at (-2.07, 1.1);
      \vertex (b) at (-2.07, -1.1);
      \vertex (v1) at (-0.6,0);
      \vertex[blob,scale=0.8] (vm) at (0.45,0) {};
      \vertex (v2) at (1.5,0);
      \vertex (c) at (2.97,1.1);
      \vertex (d) at (2.97,-1.1);
      \diagram* {
        (a) --[fermion, edge label=$e^-$] (v1),
        (b) --[anti fermion, edge label'=$e^+$] (v1),
        (v1) --[photon, edge label={$\gamma,Z$}] (vm),
        (vm) --[photon, edge label={$\gamma,Z$}] (v2),
        (v2)--[fermion,edge label=$\mu^-$] (c),
        (v2)--[anti fermion,edge label'=$\mu^+$] (d),
        };
    \end{feynman}
\end{tikzpicture}
\hfill
\begin{tikzpicture}[scale=0.75]
    \begin{feynman}
      \vertex (a) at (-2.07, 1.1);
      \vertex (b) at (-2.07, -1.1);
      \vertex (v1) at (-0.6,0);
      \vertex[blob, scale=0.8] (vm) at (0.45,0) {};
      \vertex (v2) at (1.5,0);
      \vertex (c) at (2.97,1.1);
      \vertex (d) at (2.97,-1.1);
      \diagram* {
        (a) --[fermion, edge label=$e^-$] (v1),
        (b) --[anti fermion, edge label'=$e^+$] (v1),
        (v1) --[photon, edge label={$\gamma,Z$}] (vm),
        (vm) --[photon, edge label={$Z,\gamma$}] (v2),
        (v2)--[fermion,edge label=$\mu^-$] (c),
        (v2)--[anti fermion,edge label'=$\mu^+$] (d),
        };
    \end{feynman}
\end{tikzpicture}
\hfill
\begin{tikzpicture}[scale=0.73]
    \begin{feynman}
      \vertex (a) at (-2.07, 1.1);
      \vertex (b) at (-2.07, -1.1);
      \vertex[blob,scale=0.8] (v1) at (-0.6,0) {};
      \vertex (v2) at (1.5,0);
      \vertex (c) at (2.97,1.1);
      \vertex (d) at (2.97,-1.1);
      \diagram* {
        (a) --[fermion, edge label=$e^-$] (v1),
        (b) --[anti fermion, edge label'=$e^+$] (v1),
        (v1) --[photon, edge label={$\gamma,Z$}] (v2),
        (v2)--[fermion,edge label=$\mu^-$] (c),
        (v2)--[anti fermion,edge label'=$\mu^+$] (d),
        };
    \end{feynman}
\end{tikzpicture}
\hfill
\begin{tikzpicture}[scale=0.73]
    \begin{feynman}
      \vertex (a) at (-2.07, 1.1);
      \vertex (b) at (-2.07, -1.1);
      \vertex (v1) at (-0.6,0);
      \vertex[blob,scale=0.8] (v2) at (1.5,0) {};
      \vertex (c) at (2.97,1.1);
      \vertex (d) at (2.97,-1.1);
      \diagram* {
        (a) --[fermion, edge label=$e^-$] (v1),
        (b) --[anti fermion, edge label'=$e^+$] (v1),
        (v1) --[photon, edge label={$\gamma,Z$}] (v2),
        (v2)--[fermion,edge label=$\mu^-$] (c),
        (v2)--[anti fermion,edge label'=$\mu^+$] (d),
        };
    \end{feynman}
\end{tikzpicture}

\hfill

\begin{tikzpicture}[scale=0.90]
    \begin{feynman}
      \vertex (a) at (0,0);
      \vertex (v1) at (1.3,0);
      \vertex (v2) at (2.6,0);
      \vertex (b) at (4,0);
      \vertex (v3) at (1.3,1.5  );
      \vertex (v4) at (2.6,1.5);
      \vertex (c) at (0,1.5);
      \vertex (d) at (4,1.5);
      
      \diagram* {
        (a) --[anti fermion,edge label'=$e^+$] (v1),
        (v1)--[boson, edge label'=$Z$] (v2),
        (v2)--[anti fermion, edge label'=$\mu^+$] (b),
        (v1) --[anti fermion, edge label=$e$] (v3),
        (v2)--[fermion ,edge label'={{$\mu$}}] (v4),
        (v3)--[boson,edge label={$Z$}] (v4),
        (c) --[fermion, edge label=$e^-$] (v3),
        (v4) --[fermion, edge label=$\mu^-$] (d),        
        };
    \end{feynman}
\end{tikzpicture}
\hfill
\begin{tikzpicture}[scale=0.90]
    \begin{feynman}
      \vertex (a) at (0,0);
      \vertex (v1) at (1.3,0);
      \vertex (v2) at (2.6,0);
      \vertex (b) at (4,0);
      \vertex (v3) at (1.3,1.5  );
      \vertex (v4) at (2.6,1.5);
      \vertex (c) at (0,1.5);
      \vertex (d) at (4,1.5);
      
      \diagram* {
        (a) --[anti fermion,edge label'=$e^+$] (v1),
        (v1)--[boson] (v4),
        (v2)--[anti fermion, edge label'=$\mu^+$] (b),
        (v1) --[anti fermion, edge label=$e$] (v3),
        (v2)--[fermion ,edge label'={{$\mu$}}] (v4),
        (v3)--[boson] (v2),
        (c) --[fermion, edge label=$e^-$] (v3),
        (v4) --[fermion, edge label=$\mu^-$] (d),        
        };
    \end{feynman}
    \node at (2,1.4) {$Z$};
    \node at (2,0.1) {$Z$};
\end{tikzpicture}
\hfill
\begin{tikzpicture}[scale=0.90]
    \begin{feynman}
      \vertex (a) at (0,0);
      \vertex (v1) at (1.3,0);
      \vertex (v2) at (2.6,0);
      \vertex (b) at (4,0);
      \vertex (v3) at (1.3,1.5  );
      \vertex (v4) at (2.6,1.5);
      \vertex (c) at (0,1.5);
      \vertex (d) at (4,1.5);
      
      \diagram* {
        (a) --[anti fermion,edge label'=$e^+$] (v1),
        (v1)--[boson, edge label'=$W$] (v2),
        (v2)--[anti fermion, edge label'=$\mu^+$] (b),
        (v1) --[anti fermion, edge label=$\nu_e$] (v3),
        (v2)--[fermion ,edge label'={{$\nu_\mu$}}] (v4),
        (v3)--[boson,edge label={$W$}] (v4),
        (c) --[fermion, edge label=$e^-$] (v3),
        (v4) --[fermion, edge label=$\mu^-$] (d),        
        };
    \end{feynman}
\end{tikzpicture}
\hfill
\begin{tikzpicture}[scale=0.90]
    \begin{feynman}
      \vertex (a) at (0,0);
      \vertex (v1) at (1.3,0);
      \vertex (v2) at (2.6,0);
      \vertex (b) at (4,0);
      \vertex (v3) at (1.3,1.5  );
      \vertex (v4) at (2.6,1.5);
      \vertex (c) at (0,1.5);
      \vertex (d) at (4,1.5);
      
      \diagram* {
        (a) --[anti fermion,edge label'=$e^+$] (v1),
        (v1)--[boson] (v4),
        (v2)--[anti fermion, edge label'=$\mu^+$] (b),
        (v1) --[anti fermion, edge label=$\nu_e$] (v3),
        (v2)--[fermion ,edge label'={{$\nu_\mu$}}] (v4),
        (v3)--[boson] (v2),
        (c) --[fermion, edge label=$e^-$] (v3),
        (v4) --[fermion, edge label=$\mu^-$] (d),        
        };
    \end{feynman}
    \node at (2,1.4) {$W$};
    \node at (2,0.1) {$W$};
\end{tikzpicture}

\caption{Weak-type diagrams contributing to the $e^+e^-\to \mu^+\mu^-$ process.}
\label{fig:WEAK_ee_mm}
\end{figure}

\begin{figure}
\hfill
\begin{tikzpicture}[scale=0.90]
    \begin{feynman}
      \vertex (a) at (-0.6,0);
      \vertex[blob,scale=0.8] (am) at (0.45,0) {};
      \vertex (v1) at (1.5,0);
      \vertex (b) at (3.17,1.3);
      \vertex (c) at (3.17,-1.3);
      \diagram* {
        (a) --[photon, edge label={$\gamma,Z$}] (am),
        (am) --[photon, edge label={$\gamma,Z$}] (v1),
        (v1)--[fermion,edge label=$\mu^-$] (b),
        (v1)--[anti fermion,edge label'=$\mu^+$] (c)
        };
    \end{feynman}
\end{tikzpicture}
\hfill
\begin{tikzpicture}[scale=0.90]
    \begin{feynman}
      \vertex (a) at (-0.6,0);
      \vertex[blob,scale=0.8] (am) at (0.45,0) {};
      \vertex (v1) at (1.5,0);
      \vertex (b) at (3.17,1.3);
      \vertex (c) at (3.17,-1.3);
      \diagram* {
        (a) --[photon, edge label={$\gamma,Z$}] (am),
        (am) --[photon, edge label={$Z,\gamma$}] (v1),
        (v1)--[fermion,edge label=$\mu^-$] (b),
        (v1)--[anti fermion,edge label'=$\mu^+$] (c)
        };
    \end{feynman}
\end{tikzpicture}
\hfill
\begin{tikzpicture}[scale=0.90]
    \begin{feynman}
      \vertex (a) at (-0.6,0);
      \vertex[blob,scale=0.8] (v1) at (1.5,0) {};
      \vertex (b) at (3.17,1.3);
      \vertex (c) at (3.17,-1.3);
      \diagram* {
        (a) --[photon, edge label={$\gamma,Z$}] (v1),
        (v1)--[fermion,edge label=$\mu^-$] (b),
        (v1)--[anti fermion,edge label'=$\mu^+$] (c)
        };
    \end{feynman}
\end{tikzpicture}
\hfill
\hfill

\caption{Weak-type diagrams contributing to the $Z\to \mu^+\mu^-$ process.}
\label{fig:WEAK_Z_mm}
\end{figure}
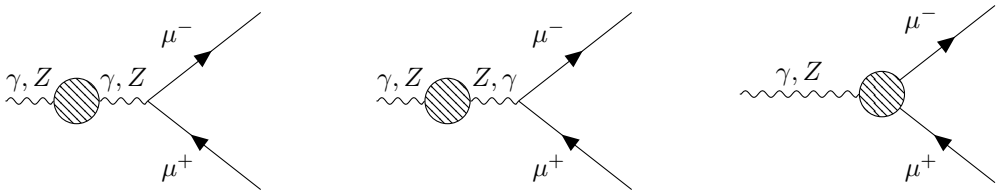

When calculated at the $Z$ pole, the radiative corrections to $Z\to \mu^+ \mu^-$ are linked to the radiative corrections to the process $ e^+ e^- \to  \mu^+ \mu^-$ at one loop level. This allows for the use of the measurements performed at LEP of the latter to constrain the higher-order contributions of new physics processes to the former.
In all calculations of one-loop processes, we will use the on-shell renormalization procedure developed in~\cite{Hollik:1988ii,Bohm:1986rj}. This scheme conveniently uses parameters with clear physical meaning that can be measured experimentally and also allows for the separation of QED, and Weak corrections into two distinct gauge invariant classes. In this classification, QED corrections correspond to diagrams with a virtual photon exchanged in the loop and diagrams with real photon emission in the form of \textit{bremsstrahlung} radiation originating from the initial or final state particles. The one-loop QED-type diagrams contributing to the processes $ e^+ e^- \to  \mu^+ \mu^-$ and $Z\to \mu^+ \mu^-$ are represented in Figures \ref{fig:QED_ee_mm} and \ref{fig:QED_Z_mm}, respectively. Summing over these diagrams eliminates IR divergences and the result is UV finite by itself after renormalization \cite{Albert:1979ix}. The remaining one-loop diagrams correspond to the Weak corrections and include the corrections to the vertices, corrections to the photon and $Z$ propagators, and box diagrams with massive boson exchange. These are represented for each process in Figs. \ref{fig:WEAK_ee_mm} and \ref{fig:WEAK_Z_mm}. The QED corrections depend on the experimental setup, whereas the Weak corrections depend only on the underlying structure of the theory. In the case of colored final states, additional QCD corrections have to be included. These are formally identical to the QED corrections \cite{Albert:1979ix}.

As previously stated, partial widths can only be compared with experiment once the effects of final-state radiation are included. In the case of the $Z$ decaying into leptons ($l$), the final state QED corrections are accounted for by introducing a multiplicative factor as
\begin{equation}
    \Gamma (Z\to  l^+ l^- )=\Gamma^{0} (Z\to l^+ l^- )F_{\textup{QED}}^l,
\end{equation}
\noindent where, $F_ {\textup{QED}}^l=1+\frac{3}{4}Q_l^2\frac{\alpha(m_Z^2)}{\pi}+\mathcal{O}\left(\alpha^2\right)$ \cite{Albert:1979ix}. At leading order in $\alpha$ this factor is independent of the lepton species. As a consequence, these corrections effectively cancel out in the measured $R_{e\mu}$ value which, in turn, allows one to identify $R_{e\mu} \approx R_{e\mu}^0$.

The weak corrections are expected to change the chiral structure of the theory at one loop level. However, the matrix elements that do not preserve the tree-level structure can be neglected \cite{Bohm:1986rj}. The propagator corrections at one loop level come from the mixing of the photon and $Z$ propagators and from the $Z$ and photon exchange diagrams. These enter as redefinitions of the QED and neutral current coupling strengths, $Z$ width, and sine of the Weinberg angle. The vertex corrections are defined as the $Z \mu^+ \mu^-$ three-point functions at one loop level and can be interpreted as additive contributions to the left and right-handed coupling constants. In the on-shell renormalization scheme, the counterterms needed to renormalize the vertex depend on the bare self-energy of the mixed $\gamma-Z$ propagator and on the external legs' self-energy \cite{Hollik:1993cg}.

\subsection{New Physics Contributions}

In order to be able to constrain the parameters stemming from the new physics Lagrangian, it is important to divide the different pieces of the one-loop corrections into the two and three-point function related to the $Z \mu^+ \mu^-$ vertex.    

One-loop corrections to the bosonic propagator are expected in the presence of new physics. These can be absorbed into the definition of the effective Weinberg angle and are parameterized by the Peskin-Takeuchi $S$, $T$, and $U$ oblique parameters \cite{Peskin:1991sw}. The deviation from the SM prediction can be written in the form~\cite{Grant:1998kp}
%
%
\begin{equation}
    \delta s_W^2=s_W^{\text{eff}\,2}-s_W^{\text{SM}\,2}=\frac{\alpha}{c_W^2-s_W^2}\left(\frac{1}{4}S-s_W^2 c_W^2T\right),
\end{equation}
where $s_W^{\text{eff}}$ is the sine of the effective Weinberg angle, which accounts for all known and unknown corrections, and $s_W^{\text{SM}}$ is the sine of the SM effective Weinberg angle. In the following discussion, we use the experimental value $s_W^2=[s_W^{\text{eff 2}}]_{\textup{EXP}}\approx 0.23148$ \cite{ParticleDataGroup:2022pth}. The most recent experimental data shows the oblique parameters agreeing with the SM value of zero \cite{ParticleDataGroup:2022pth}. In both models studied in Section \ref{sec:results}, we follow the procedure outlined in \cite{Capucha:2022kwo} to constrain our parameter space accordingly and neglect new physics' direct contributions to the propagator. We do, however, take into account how new physics affects the $T$ parameter, which is the most relevant to constrain our parameter space, but do so in the scan prior to computing the corrections to the muon-related observables. Additional contributions coming from box diagrams would have to be included, but neither model we chose to work with has new fields coupling to the first-generation fermions. 
It is important to note that whatever the model under study is, since we can write the corrections as a function of parameters that are very constrained, the oblique parameters, these corrections will always be negligible.


New physics corrections to the vertex are absorbed into a redefinition of the left and right $Z \mu^+ \mu^-$ couplings. At one-loop, the renormalized couplings can be spit as $\hat{g}_{L,R}=g_{L,R}^{\textup{SM}}+\delta \widetilde{g}_{L,R}$, where the first term contains the tree-level and all SM one-loop contributions, and $\delta \widetilde{g}_{L,R}$ all one loop vertex contributions from new physics. From now on we drop the muon lower index since there will be no source of confusion about which lepton we are referring to. The new physics term is defined as \cite{Hollik:1993cg}
\begin{equation}
    \delta \widetilde{g}_{L,R}=-\frac{c_W s_W}{e}\widetilde{\Gamma}^{Z\mu\mu}_{L,R}+g_{L,R}\delta \widetilde{Z}_{L,R}-g_{L,R}\frac{c_W}{s_W}\frac{\widetilde{\Sigma}^{\gamma Z}(0)}{m_Z^2}  +s_Wc_W \frac{\widetilde{\Sigma}^{\gamma Z}(0)}{m_Z^2}.
    \label{eq:NP_correction_g}
\end{equation}
where a $+\delta s_W^2$ factor would have to be added to the effective coupling if the oblique parameters were not neglected \cite{PhysRevD.62.015011}. The first term in \eqref{eq:NP_correction_g} is computed from the bare three-point function containing the new particles of the model. The decomposition into left and right components is done as\footnote{Where we neglect terms proportional to the anomalous weak-electric and magnetic dipole moments (\textit{i.e.} $\propto \sigma_{\mu \nu}k^\nu$, with $k^\nu$ the momentum of the $Z$ boson) because of a $m_\mu / m_Z$ suppression when compared to $\widetilde{\Gamma}_{L,R}^{Z\mu\mu}$ \cite{Bernabeu:1995sq}.}
\begin{equation}
   i \widetilde{\Gamma}_\lambda^{Z\mu\mu}=
   \vcenter{\hbox{\begin{tikzpicture}
    \begin{feynman}[every blob={/tikz/fill=gray!30,/tikz/inner sep=2pt}, scale=0.75,transform shape]
      \vertex (a) at (-0.6,0);
      \vertex[blob] (v1) at (1.5,0) {};
      \vertex (b) at (3.17,1.3);
      \vertex (c) at (3.17,-1.3);
      \diagram* {
        (a) --[photon, edge label={$Z$}] (v1),
        (v1)--[fermion,edge label=$\mu^-$] (b),
        (v1)--[anti fermion,edge label'=$\mu^+$] (c)
        };
    \end{feynman}
\end{tikzpicture}}} \approx\ \  -\frac{i e}{s_W c_W} \gamma_\lambda \left( \widetilde{\Gamma}_L^{Z\mu\mu} P_L  +  \widetilde{\Gamma}_R^{Z\mu\mu} P_R \right).
\end{equation}
Furthermore, $\widetilde{\Sigma}^{\gamma Z}$ can be calculated from the sum of diagrams containing new particles that contribute to the mixed $\gamma-Z$ bare self-energy\footnote{The longitudinal term proportional to $k^{\alpha}k^{\beta}$ can be neglected because it is suppressed by a $m_f^2/m_Z^2$ factor \cite{Hollik:1993cg}.}
\begin{equation}
 - i \widetilde{\Sigma}^{\gamma Z}(k^2) g^{\alpha \beta} = 
 \vcenter{\hbox{\begin{tikzpicture}
    \begin{feynman}[every blob={/tikz/fill=gray!30,/tikz/inner sep=2pt}]
      \vertex (a) at (-1.5, 0);
      \vertex (b) at (1.5, 0);
      \vertex[blob] (v1) at (0,0) {};
      \diagram* {
        (a) --[boson, momentum={$k$}, edge label'=$\gamma$] (v1),
        (v1) --[boson, momentum={$k$}, edge label'=$Z$] (b)
        };
    \end{feynman}
\end{tikzpicture}}}
\end{equation}
calculated at zero momentum transfer $k^2=0$. Finally, $\delta \widetilde{Z}_{L,R}$ is the counter-term for the left or right-handed component of the muon wave function,
\begin{equation}
    \delta \widetilde{Z}_{L,R}=-\widetilde{\Sigma}_{L,R}(m_\mu^2)-m_\mu^2\left[ \widetilde{\Sigma}'_L(m_\mu^2) +\widetilde{\Sigma}'_R(m_\mu^2) + 2\widetilde{\Sigma}'_S(m_\mu^2)\right],
\end{equation}
where the functions $\widetilde{\Sigma}_{L,R,S}$ come from the decomposition of the muon bare self-energy into

\begin{equation}
    i\widetilde{\Sigma}_\mu(\slashed{p}) =
    \vcenter{\hbox{\begin{tikzpicture}
    \begin{feynman}[every blob={/tikz/fill=gray!30,/tikz/inner sep=2pt}]
      \vertex (a) at (-1.5, 0);
      \vertex (b) at (1.5, 0);
      \vertex[blob] (v1) at (0,0) {};
      \diagram* {
        (a) --[fermion, momentum={$p$}, edge label'=$\mu^-$] (v1),
        (v1) --[fermion, momentum={$p$}, edge label'=$\mu^-$] (b)
        };
    \end{feynman}
\end{tikzpicture}}}
 = i\slashed{p}\widetilde{\Sigma}_L(p^2)P_L + i\slashed{p}\widetilde{\Sigma}_R(p^2)P_R + im_\mu \widetilde{\Sigma}_S(p^2).
\end{equation}
with $\Sigma'=\frac{\partial \Sigma}{\partial s}$.

\subsection{Connection with experiment}

As shown, the most significant new physics contributions come from the $Z \mu^+ \mu^-$ vertex corrections. To a good approximation, $\mathcal{A}_\mu^0$ and $R_{e\mu}$ can be defined at one-loop order by replacing the tree-level couplings in \eqref{eq:Amu} and \eqref{eq:Rem} by $\hat{g}_{L,R}=g_{L,R}^{\textup{SM}}+\delta \widetilde{g}_{L,R}$. In the presence of new physics, we expect the theoretical prediction of the observables $\mathcal{A}_\mu^0$ and $R_{e\mu}$ to deviate from the SM. This deviation can be written as
\begin{subequations}
    \begin{equation}
       \mathcal{A}_\mu^0=\mathcal{A}_\mu^{\textsc{SM}}+\delta \widetilde{\mathcal{A}}_\mu,
    \end{equation}
    \begin{equation}
        R_{e\mu} = R_{e\mu}^{\textsc{SM}}+\delta \widetilde{R}_{e\mu}.
    \end{equation}
\end{subequations}
The relation between the new physics contributions to the couplings and the deviations in the observables can be derived by taking $|\delta \widetilde{g}_{L,R}| \ll |g_{L,R}^{\textup{SM}}|$, and expanding $\mathcal{A}_\mu^0$ and $R_{e\mu}$ to first order in $\delta \widetilde{g}_{L,R}$. 
In the small mass limit ($\mu_\mu \to 0$) we obtain
\begin{subequations}
    \begin{equation}
       \delta \widetilde{\mathcal{A}}_\mu\approx -3.6343\delta \widetilde{g}_L-4.2154\delta \widetilde{g}_R ,
    \end{equation}
    \begin{equation}
        \delta \widetilde{R}_{e\mu}\approx -4.2619\delta \widetilde{g}_L+3.6744\delta \widetilde{g}_R,
    \end{equation}
    \label{dObs_dg}
\end{subequations}

Using this formalism, the parameter space of models with new couplings to the muon can be systematically constrained, by requiring the deviation from the SM predictions to fall within the experimental uncertainties presented in Table \ref{table:compare_SM}.

\section{Results \label{sec:results}}

In order to see how these observables can be used to test and constrain extensions of the SM, we consider the models presented in~\cite{PhysRevD.102.075009, Capucha:2022kwo}. We perform a multi-parameter scan to identify the allowed parameter space obeying the already implemented flavor constraints \textit{and} the newly defined conditions arising from the muon-related observables. The flavor-related constraints are the new $R\left(K^{\left(*\right)}\right)$ now in agreement with the SM predictions (as discussed in the introduction), $\mathcal{B}\left(B_S\to\mu^+\mu^-\right)$, $B_S-\Bar{B}_S$ mixing and $b\to s\gamma$. 
We force both models to be able to explain the $g-2$ anomaly which gives a special role to the muon couplings. Besides that, we also force the DM candidate to saturate the relic density measurement, take into account the bounds from direct and indirect detection, 
and consider the constraints imposed by collider searches (see~\cite{PhysRevD.102.075009, Capucha:2022kwo} for a detailed discussion).

The models add two scalar fields, $\Phi_2$ and $\Phi_3$ (the second with a color charge), and a vector-like fermion $\chi$. These fields are collectively called the \textit{Dark Sector}. A new $Z_2$ symmetry is introduced, under which the SM fields are even and the Dark Sector fields are odd. In the one we call Model 3, the new fermion is a singlet of $SU(2)_L$ and the scalars are doublets. In Model 5, the fermion is an $SU(2)_L$ doublet and the scalars are singlets. In both models, the neutral component of the $\Phi_2$ scalar is split into its real and imaginary degrees of freedom ($\frac{1}{\sqrt{2}}(S+iA)$). The flavor phenomenology is the same for both $S$ and $A$, hence, without loss of generality, we assume $m_S<m_A$, effectively choosing $S$ to be the DM candidate. The $SU(3)_c$ and $SU(2)_L$ representations and weak hypercharges of the Dark Sector particles of both models are summed up in Table \ref{table:M3_M5_Qnt_numbers}.

\begin{table}[h!]
\centering
\hfill
\begin{tabular}{c c c c}
\multicolumn{4}{c}{Model 5}\\
\hline\hline
  & $SU(3)_c$ & $SU(2)_L$ & $U(1)_Y$ \\
\hline
$\chi_R$	&	\textbf{1}	&	\textbf{2}	 &	$-1/2$ \\
$\Phi_2$	&	\textbf{1}	&	\textbf{1}	 &	$0$ \\
$\Phi_3$	&	\textbf{3}	&	\textbf{1}	 &	$+2/3$ \\
\hline
\end{tabular}
\hfill
\begin{tabular}{c c c c}
\multicolumn{4}{c}{Model 3}\\
\hline\hline
  & $SU(3)_c$ & $SU(2)_L$ & $U(1)_Y$ \\
\hline
$\chi_R$	&	\textbf{1}	&	\textbf{1}	 &	$-1$ \\
$\Phi_2$	&	\textbf{1}	&	\textbf{2}	 &	$+1/2$ \\
$\Phi_3$	&	\textbf{3}	&	\textbf{2}	 &	$+7/6$ \\
\hline
\end{tabular}
\hfill
\caption{Representations and weak hypercharges of Dark Sector of Model 5 (left) and Model 3 (right). Adapted from \cite{PhysRevD.102.075009} and \cite{Capucha:2022kwo}.}

\label{table:M3_M5_Qnt_numbers}
\end{table}

LEP searches for unstable heavy vector-like charged leptons set a lower bound on the mass of $\chi^\pm$ of 101.2 GeV \cite{L3:2001xsz}. Other more recent constraints from the LHC do not apply, since these assume either that the vector-like lepton couples to the tau \cite{CMS:2019emo} or a very small amount of missing transverse energy in the final states \cite{Bissmann:2020lge}. Other limits on the masses come from the flavor constraints and are detailed in \cite{PhysRevD.102.075009, Capucha:2022kwo}.

In the following two sections, we present the new physics contributions to $\mathcal{A}_\mu^0$ and $R_{e\mu}$, and the results of the scans for the two models. We performed the random scan and checked which conditions were satisfied for each point. The relic density was computed with {\fontfamily{lmtt}\selectfont micrOMEGAs} \cite{Belanger:2020gnr}, which solves the Boltzmann equations numerically. We show projections from the parameter space with a similar color scheme to that used in \cite{PhysRevD.102.075009, Capucha:2022kwo}, where all \textbf{cyan} points satisfy the mass constraints and explain the $B$ meson data (including the latest reinterpretation of the LHCb collaboration) in the 2$\sigma$ range. In addition to the previous constraints, the \textbf{blue} points agree with DM relic abundance observations in the 2$\sigma$ range. Similarly, the \textbf{green} points verify all previous constraints, together with the restrictions coming from DM direct detection and collider searches. Finally, the \textbf{red} points satisfy all previous constraints and agree with the muon $g-2$ experiments within $3\sigma$. We updated experimental values used in these conditions in light of recent results. In particular, the more accurate analysis done by the LHCb Collaboration of the exclusive branching fractions, $R(K)$ and $R(K^*)$, \cite{LHCb:2022qnv} has implications in the $C_9^{NP}$ Wilson coefficient used in \cite{PhysRevD.102.075009}. We constrained all parameter space points to be within the $2\sigma$ range around the new best-fitted value $C_9^{NP} = -0.19 \pm 0.06 $ \cite{Alguero:2023jeh}. Furthermore, bounds on Dark Matter relic density are from the 2018 Planck Collaboration \cite{ planck}, and we have included the most recent upper bounds on the spin-independent DM-nucleon scattering cross-section from the LZ experiment \cite{LZnew}.


For every point that verified all previous conditions, we computed $\delta \widetilde{\mathcal{A}}_\mu$ and $\delta \widetilde{R}_{e\mu}$ and checked whether $\mathcal{A}_\mu^{\textup{SM}}+\delta \widetilde{\mathcal{A}}_\mu$ and $R_{e\mu}^{\textup{SM}}+\delta \widetilde{R}_{e\mu}$ were kept within their $1\sigma$ experimental confidence levels presented in Table \ref{table:compare_SM}, that is
\begin{subequations}
    \begin{equation}
       -0.0203<\delta\widetilde{\mathcal{A}}_\mu<0.0097,
    \end{equation}
    \begin{equation}
       -0.0020 <\delta\widetilde{R}_{e\mu} <0.0038.
    \end{equation}
\end{subequations}

\noindent In the discussion that follows, parameter space points that verified this condition were assigned the color \textbf{yellow}.

As noted in \cite{Capucha:2022kwo}, the scalar fields in Model 3 introduce a nonzero contribution to the oblique parameter $T$, which is given by \cite{Grimus:2007if}
\begin{equation}
    T=\frac{g^2}{64\pi^2 m_W^2\alpha}\left[F(m_{\phi_l}^2,m_S^2)+F(m_{\phi_l}^2,m_A^2)-F(m_S^2,m_A^2)\right],
\end{equation}
where $g$ is the $SU(2)_L$ coupling constant, $m_W$ is the mass of the $W$ boson and $\alpha$ the fine-structure constant. The function $F(A,B)$ is defined as
\begin{equation}\label{eq:FAB}
    F(A,B)=\begin{cases}
      \frac{A+B}{2}-\frac{AB}{A-B}\log\frac{A}{B}, \text{ if $A\neq B$.}\\
      0\hspace{2.61cm},\text{ if $A=B$.}
    \end{cases}
\end{equation}
The vector-like fermion and the colored scalars have a vanishing contribution to the $T$ parameter. The first contributes through a diagram similar to the photon self-energy in QED, which at zero momentum transfer vanishes, and the colored scalars produce a term proportional to $F(m_{\phi_q^{5/3}}^2,m_{\phi_q^{2/3}}^2)$, which is zero by \eqref{eq:FAB} since we consider $m_{\phi_q^{5/3}}=m_{\phi_q^{2/3}}$. To account for the effect of new physics in the oblique parameter $T$, we impose a  supplementary condition that every point in the Model 3 scan must obey, which is to be within the 2$\sigma$ experimental uncertainty interval. At present, the best experimental value for this quantity is $T=0.03 \pm 0.12$ \cite{ParticleDataGroup:2022pth}. In the case of Model 5, the contribution to the $T$ parameter is zero \cite{PhysRevD.102.075009} and it provides no constraint on the model's parameters. 

The models were implemented both in {\fontfamily{lmtt}\selectfont LanHEP} \cite{semenov2014lanhep} and {\fontfamily{lmtt}\selectfont FeynRules} \cite{Christensen:2008py}. We used the packages {\fontfamily{lmtt}\selectfont FeynArts} \cite{Kublbeck:1990xc} to generate the relevant diagrams and {\fontfamily{lmtt}\selectfont FeynCalc} \cite{Mertig:1990an, Shtabovenko:2016sxi} to compute the corresponding amplitudes and perform the integration. The results are written in terms of the Passarino-Veltman integral functions \cite{Passarino:1978jh}. We used the package {\fontfamily{lmtt}\selectfont LoopTools} \cite{Hahn:1998yk} to numerically evaluate the integrals.

\subsection{Model 5}

\begin{figure}[h!]
    \centering
    \includegraphics[scale=0.4]{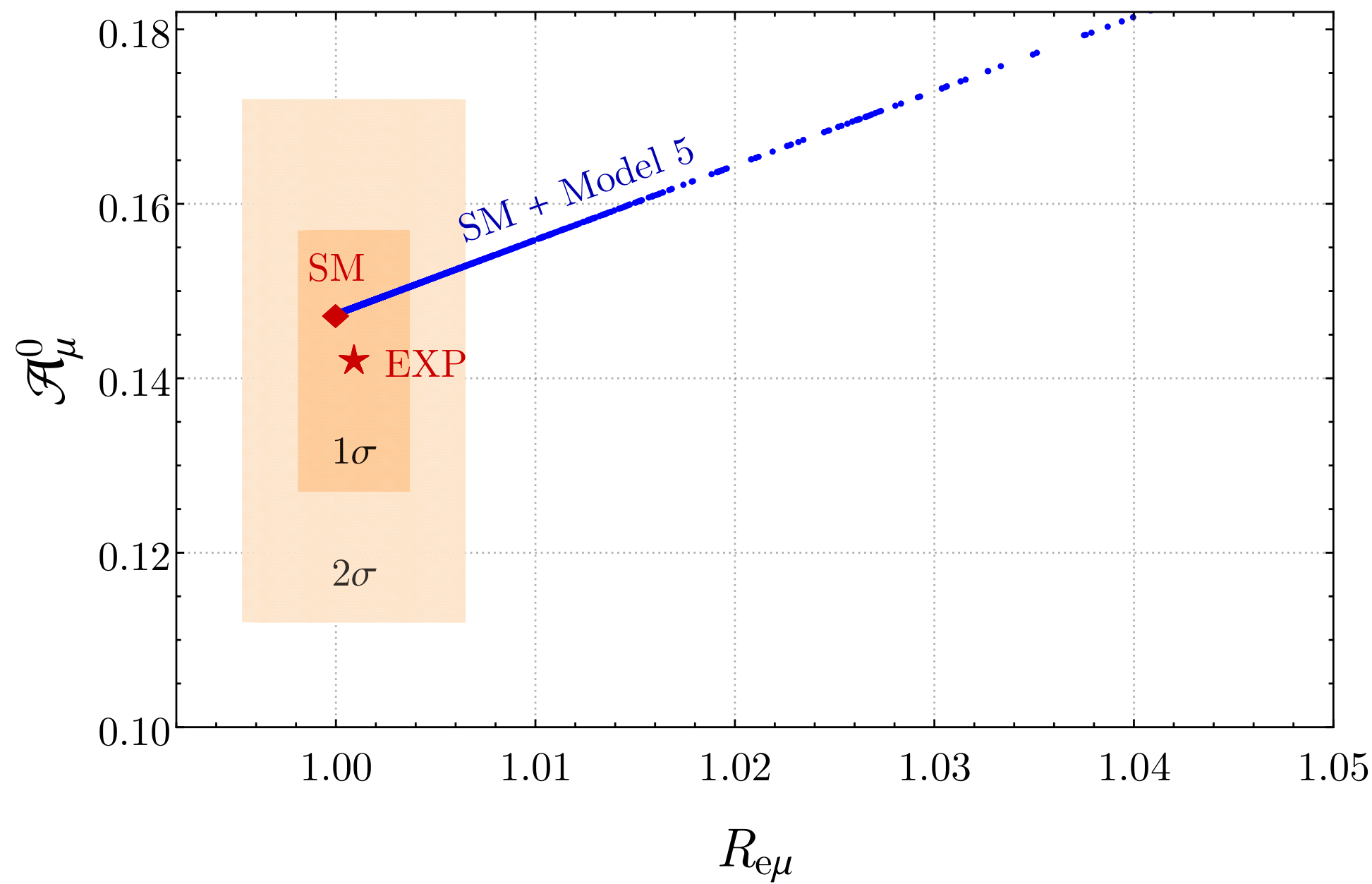}
    \caption{Representation of the $\mathcal{A}_\mu^0 - R_{e\mu}$ space. The $1\sigma$ and $2\sigma$ experimental regions are shown in different shades of orange, together with a red star marking the experimental central value. The red diamond marks the SM prediction, and the blue points show how Model 5 impacts this prediction.}
    \label{fig:mod5_Rem_Am}
\end{figure}

In Model 5, the single diagram contributing to the self-energy of the mixed $\gamma - Z$ propagator, $\Sigma^{\gamma Z}$, has the fermion $\chi$ running in the loop. When evaluated at zero momentum transfer, its amplitude was computed and confirmed to equal zero. As a consequence, the task of computing the NP corrections to the $Z \mu^+ \mu^-$ vertex was reduced down to the computation of $\widetilde{\Gamma}_\lambda^{Z\mu\mu}$ and $\widetilde{\Sigma}_\mu(\slashed{p})$. The relevant new diagrams contributing at the one-loop level are
\begin{equation}
\begin{split}
	i\widetilde{\Gamma}_\lambda^{Z\mu\mu} &= \vcenter{\hbox{ \begin{tikzpicture}
    \begin{feynman}[scale=0.90,transform shape]
      \vertex (a) at (-0.3,0);
      \vertex (v1) at (1.5,0);
      \vertex (b) at (3.17,1.3);
      \vertex (bm) at (2.33,0.65);
      \vertex (c) at (3.17,-1.3);
      \vertex (cm) at (2.33,-0.65);
      \diagram* {
        (a) --[photon, edge label={$Z$}] (v1),
        (v1)--[fermion,edge label=$\chi$] (bm),
        (bm)--[fermion,edge label=$\mu^-$] (b),
        (v1)--[anti fermion,edge label'=$\chi$] (cm),
        (cm)--[anti fermion, edge label'=$\mu^+$] (c),
        (bm)--[scalar ,edge label=$S$] (cm)
        };
    \end{feynman}
\end{tikzpicture}}} + \vcenter{\hbox{\begin{tikzpicture}
    \begin{feynman}[scale=0.90,transform shape]
      \vertex (a) at (-0.3,0);
      \vertex (v1) at (1.5,0);
      \vertex (b) at (3.17,1.3);
      \vertex (bm) at (2.33,0.65);
      \vertex (c) at (3.17,-1.3);
      \vertex (cm) at (2.33,-0.65);
      \diagram* {
        (a) --[photon, edge label={$Z$}] (v1),
        (v1)--[fermion,edge label=$\chi$] (bm),
        (bm)--[fermion,edge label=$\mu^-$] (b),
        (v1)--[anti fermion,edge label'=$\chi$] (cm),
        (cm)--[anti fermion, edge label'=$\mu^+$] (c),
        (bm)--[scalar ,edge label=$A$] (cm)
        };
    \end{feynman}
\end{tikzpicture}}} ,\\
	i\widetilde{\Sigma}_\mu(\slashed{p}) &=\vcenter{\hbox{\begin{tikzpicture}
    \begin{feynman}[scale=0.90,transform shape]
      \vertex (a) at (0,0);
      \vertex (v1) at (1.43,0);
      \vertex (v2) at (2.87,0);
      \vertex (b) at (4.3,0);
      \diagram* {
        (a) --[fermion, edge label'=$\mu^-$] (v1),
        (v1) -- [scalar, half left, looseness=1.68, edge label=$S$] (v2),
        (v2)  -- [anti fermion, half left, looseness=1.68, edge label=$\chi$] (v1),
        (v2) --[fermion, edge label'=$\mu^-$] (b)
        };
    \end{feynman}
\end{tikzpicture}}}  +\vcenter{\hbox{\begin{tikzpicture}
    \begin{feynman}[scale=0.90,transform shape]
      \vertex (a) at (0,0);
      \vertex (v1) at (1.43,0);
      \vertex (v2) at (2.87,0);
      \vertex (b) at (4.3,0);
      \diagram* {
        (a) --[fermion, edge label'=$\mu^-$] (v1),
        (v1) -- [scalar, half left, looseness=1.68, edge label=$A$] (v2),
        (v2)  -- [anti fermion, half left, looseness=1.68, edge label=$\chi$] (v1),
        (v2) --[fermion, edge label'=$\mu^-$] (b)
        };
    \end{feynman}
\end{tikzpicture}}} .
\end{split}
\end{equation}
Besides the SM parameters, the functions $\widetilde{\Gamma}_\lambda^{Z\mu\mu}$ and $\widetilde{\Sigma}_\mu(\slashed{p})$ depend only on the new masses $m_S$, $m_A$, $m_\chi$ and on the absolute value of the new Yukawa muon coupling $|y_\mu|$. We confirmed that the divergences were indeed canceled in \eqref{eq:NP_correction_g}. In addition, we found both functions to be symmetric under $m_S \leftrightarrow m_A$ and to scale as $\sim |y_\mu|^2$, as expected. In Model 5 the new particles couple to left-handed muons but not to their right-handed counterparts. As a consequence, in the limit $\mu_\mu \to 0$, Model 5 was found to only impact the left $Z \mu^+ \mu^-$ coupling, at the one-loop level. This behavior is evidenced by the blue line in Fig. \ref{fig:mod5_Rem_Am}, where the result of a random scan to the parameter space forms a 
\begin{figure}[h!]
\centering
\hfill
\includegraphics[width=0.32\textwidth]{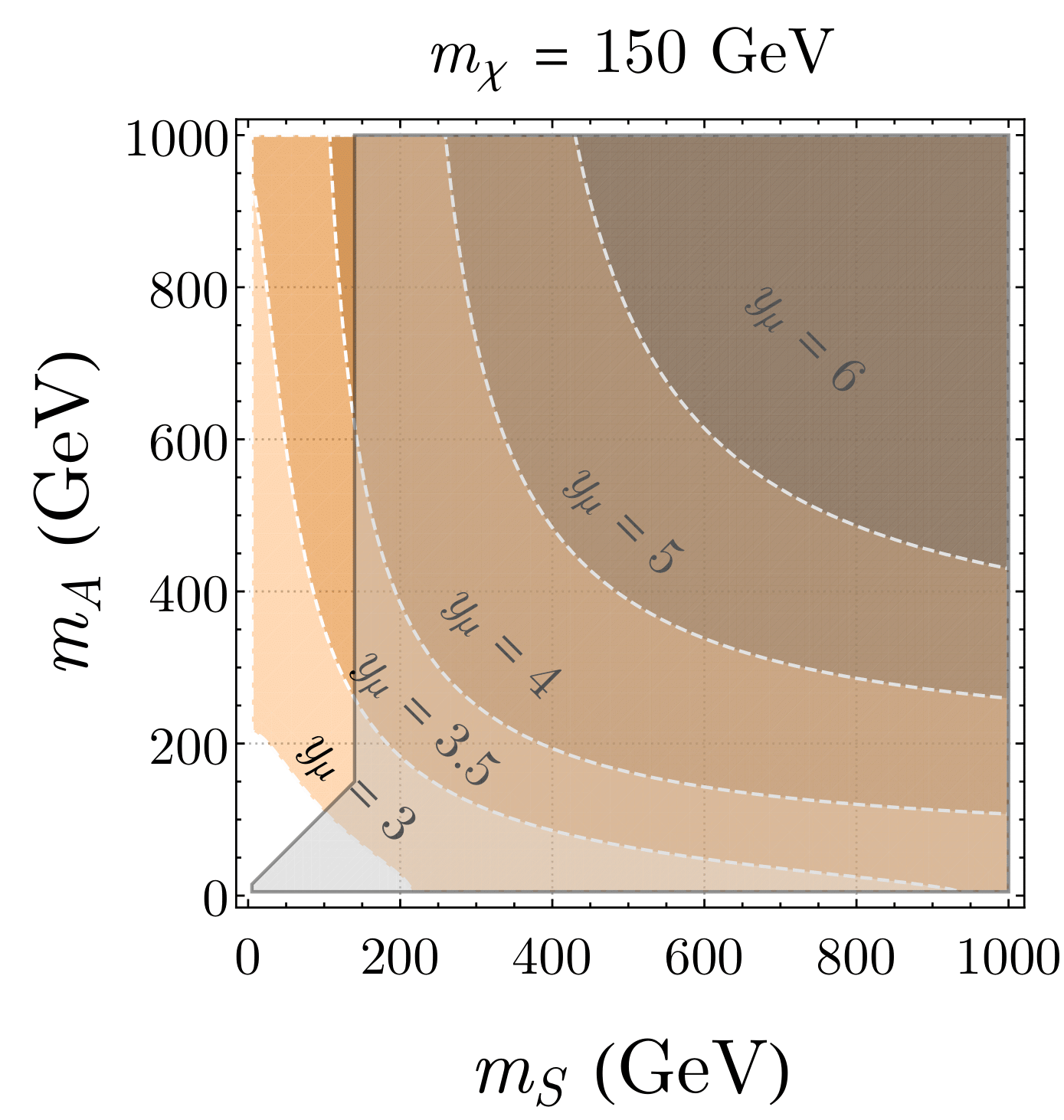}
\hfill
\includegraphics[width=0.32\textwidth]{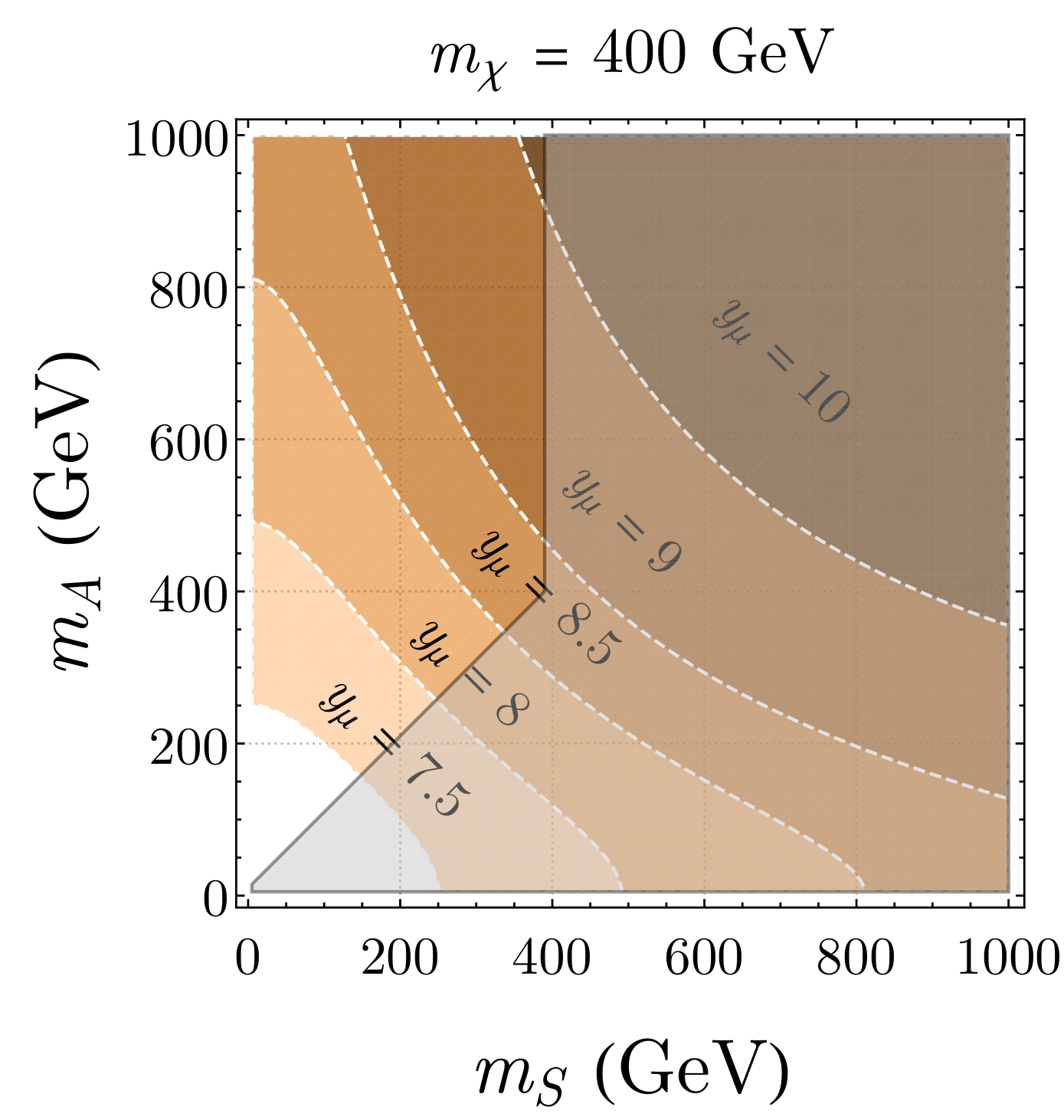}
\hfill
\includegraphics[width=0.32\textwidth]{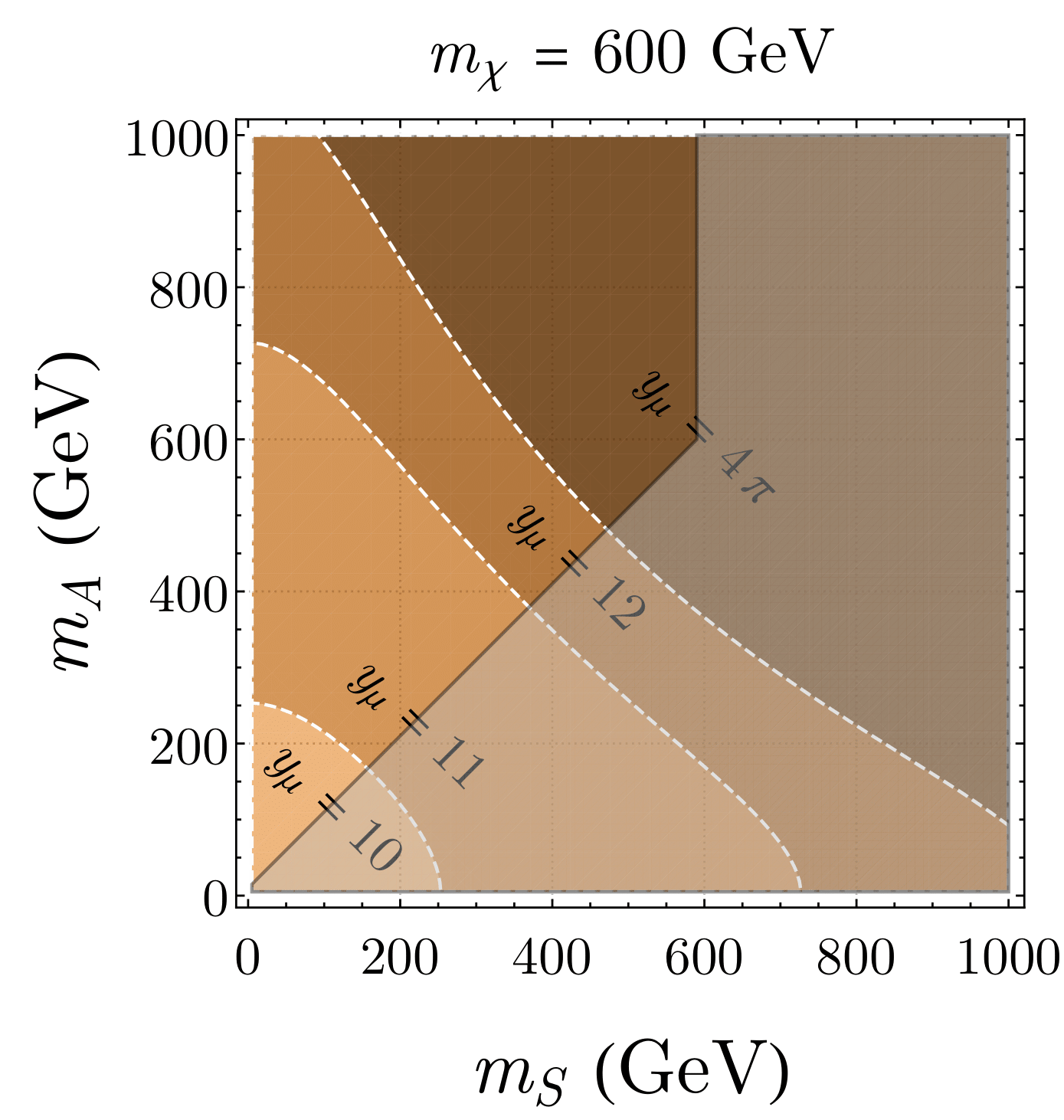}
\hfill
\caption{For $m_\chi=150$ GeV and for different values of $y_\mu$, the left panel shows the $m_S - m_A$ projection plane with the points that keep $\mathcal{A}_\mu^0$ and $R_{e\mu}$ within the $1\sigma$ uncertainty ranges. 
Darker shades of orange correspond to higher values for $y_\mu$. The middle and right panels correspond to the parameter space slices with $m_\chi=400$ GeV and $m_\chi=600$ GeV, respectively. The grey-shaded region is forbidden
because $m_S$ was chosen as the DM candidate and for scanning purposes we have chosen $m_S < m_\chi + 10$ GeV and $m_S < m_A + 10 \textup{ GeV}$ with no loss of generality.}
    \label{fig:mod5_contour}
\end{figure}
line in the $R_{e\mu} - \mathcal{A}_\mu^0$ plane. In this plot the $1\sigma$ and $2\sigma$ experimental uncertainty regions are represented in different shades of orange, the red star marks the experimental central value, and the red diamond marks the present SM prediction. The blue points show the departing from the SM prediction when $m_\chi$ was varied from 101.2 to 1000 GeV, $m_S$ from 5 GeV to 1000 GeV, $m_A$ from 15 GeV to 1000 GeV, and $y_\mu$ from 0 to $4\pi$ \cite{PhysRevD.102.075009}. It is possible to check that the new particles always give positive contributions to both $\mathcal{A}_\mu^0$ and $R_{e\mu}$ meaning that, in the domain of interest, $\delta \widetilde{g}_L \leq 0$ and $\delta \widetilde{g}_R = 0$. As a consequence of the slope of the line formed by the blue points, every point satisfying $R_{e\mu}$ at some level of confidence was found to automatically satisfy $\mathcal{A}^0_\mu$ at that same level of confidence. The fact that there are points outside the experimental uncertainty indicates that the model can be constrained using these observables.

A better understanding of the observables' impact on Model 5 is obtained by studying the dependence of $\delta \widetilde{g}_L$ on the NP parameters. In Fig. \ref{fig:mod5_contour} we plot the parameter space regions that keep both $\mathcal{A}_\mu^0$ and $R_{e\mu}$ within their $1\sigma$ experimental uncertainty range, as a function of $m_S$ and $m_A$. In the left (middle, right) panel, $m_\chi$ was fixed and set equal to $150$ ($400$, $600$) GeV. Each overlapping region has a fixed $y_\mu$, with darker shades of orange corresponding to higher coupling values. The forbidden region satisfying\footnote{The (arbitrary) choice of $+10$ GeV was originally implemented in the numerical scan to guarantee the DM candidate ($S$) to be the lightest of the Dark Sector's particles.} $m_S > m_\chi + 10$ GeV and $m_S > m_A + 10 \textup{ GeV}$ is gray-shaded. We find that higher muon coupling strengths make for more stringent constraints on the parameter space, specifically in the region of low $m_S$ and low $m_A$. Fixing $y_\mu$, we conclude that these constraints get less rigid the higher $m_\chi$ is. Besides the $m_S \leftrightarrow m_A$ symmetry being manifest, we also note that in the decoupling limit, \textit{i.e.} big masses and/or small coupling, the NP corrections behave as expected.

\begin{figure}[h!]
     \centering
     \begin{subfigure}[b]{0.44\textwidth}
         \centering
         \includegraphics[width=\textwidth]{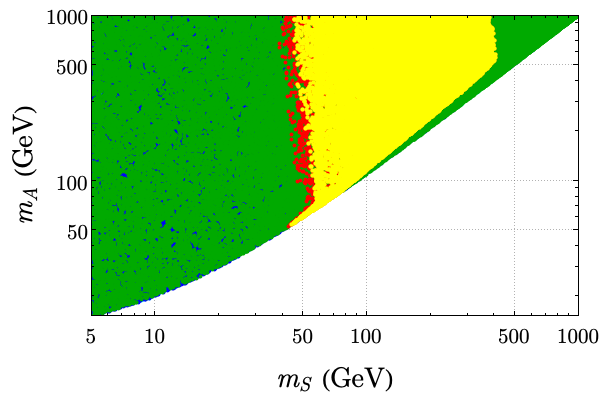}
         \caption{}
         \label{fig:mod5_a}
     \end{subfigure}
     \hfill
     \begin{subfigure}[b]{0.44\textwidth}
         \centering
         \includegraphics[width=\textwidth]{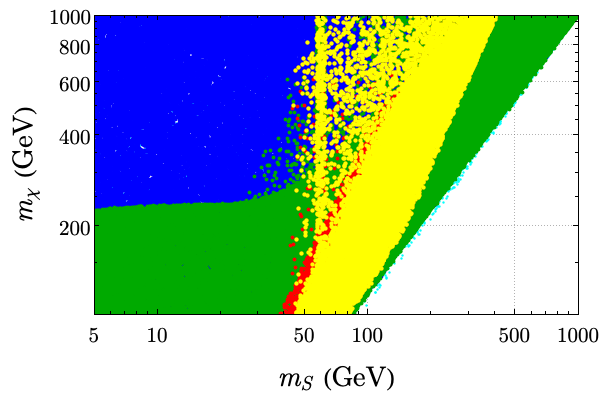}
         \caption{}
         \label{fig:mod5_b}
     \end{subfigure}
     \hfill
     \begin{subfigure}[b]{0.44\textwidth}
         \centering
         \includegraphics[width=\textwidth]{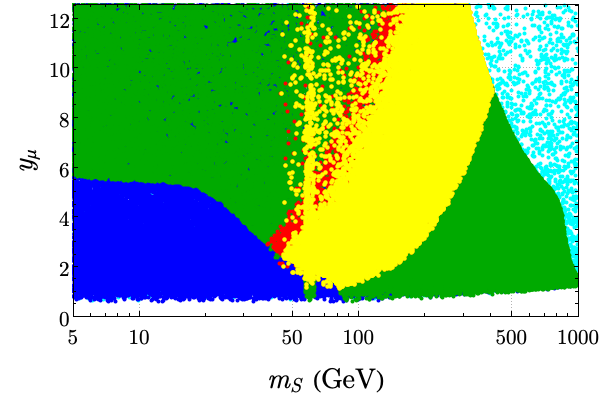}
         \caption{}
         \label{fig:mod5_c}
     \end{subfigure}
     \hfill
     \begin{subfigure}[b]{0.44\textwidth}
         \centering
         \includegraphics[width=\textwidth]{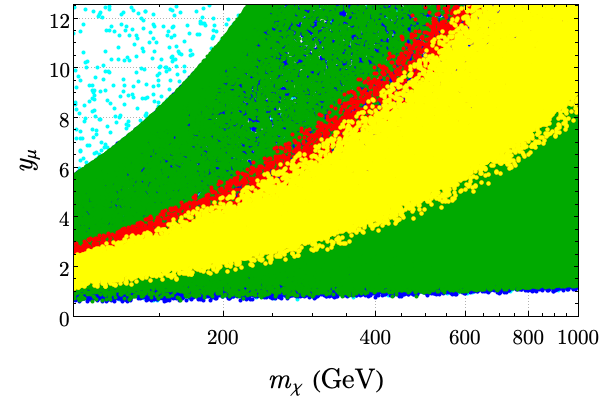}
         \caption{}
         \label{fig:mod5_d}
     \end{subfigure}     
        \caption{Allowed parameter space of Model 5 projected in the planes $m_S-m_A$ (Top Left), $m_S-m_\chi$ (Top Right), $m_S-y_\mu$ (Bottom Left), and $m_\chi-y_\mu$ (Bottom Right). The color scheme replicates the one used in \cite{PhysRevD.102.075009}. Only the yellow points agree with $\mathcal{A}_\mu^0$ and $R_{e\mu}$ at the $1\sigma$ level while satisfying all previous constraints.}
        \label{fig:mod5_all_plots}
\end{figure}

In Fig. \ref{fig:mod5_all_plots}, we show the parameter space projections relevant to the $Z \mu^+ \mu^-$ observables. The color yellow was assigned to the red data points (the ones satisfying all previous conditions) which agree with both $\mathcal{A}_\mu^0$ and $R_{e\mu}$ at the $1\sigma$ level. We have found all parameter space points to deviate from $\mathcal{A}_\mu^{\textup{SM}}$ by less than $1\sigma$, meaning all constraints arise solely due to $R_{e\mu}$. In these type of models, the $g - 2$ discrepancy can be solved by a large value of the coupling $y_\mu$.
The interplay between the high $y_\mu$ values required by $g-2$ and the low $y_\mu$ values required by the new observables is apparent in these plots. The $m_\chi - y_\mu$ projection of Fig. \ref{fig:mod5_d} is the most interesting because of the positive correlation between these parameters. In addition, it reveals a small strip excluded by $R_{e\mu}$ that was accepted by previous constraints. However, this region vanishes when we extend the threshold of the $R_{e\mu}$ requirement from $1\sigma$ to the $2\sigma$ level.

Our results (yellow points) are consistent with the previous lower bound on $m_A$ of $m_A \gtrsim 50$ GeV, but slightly disagree with the bounds on $m_S$ ($m_S \in [30, 350]$ GeV), $m_\chi$ ($m_\chi \lesssim 600$ GeV), and $y_\mu$ ($y_\mu \in [1.4,\sqrt{4\pi}] $) found in \cite{PhysRevD.102.075009}. The initial analysis lacked an observable that could impose an upper bound on $y_\mu$, and only allowed it to range up to $\sqrt{4\pi}$. The introduction of $\mathcal{A}_\mu^0$ and $R_{e\mu}$ let us push $y_\mu$ up to its perturbative limit of $4\pi$. The new region with $y_\mu > \sqrt{4\pi}$ now allows for $m_\chi$ to range from $101.2$ to $1000$ GeV unboundedly, as shown in Fig. \ref{fig:mod5_d}. Due to the positive correlation between $m_\chi$ and $m_S$ (Fig. \ref{fig:mod5_b}), the upper bound on the latter is extended to $\sim 410$ GeV, while keeping its lower bound. In sum, we argue that the bounds on the parameters are more accurately described by
\begin{align*}
&&		m_A &\gtrsim 50	\textup{ GeV},			&		y_\mu &\in [1,4\pi] , 			&&\\
&&		m_\chi & > 101.2 \textup{ GeV},		&		m_S &\in [30,410] \textup{ GeV}. &&
\end{align*}

\subsection{Model 3}

We computed the weak corrections to the left and right-handed coupling constants for Model 3. The $\gamma - Z$ mixing diagrams with the fermion $\chi$ and the scalars in the loop cancel in the zero momentum transfer limit. As such, only the vertex and muon self-energy diagrams of \eqref{eq:diagsmod3} contribute to the couplings. The fact that in this model the scalars are doublets of $SU(2)$ means that they couple to the $Z$ boson, which leads to two diagrams that were forbidden in Model 5
\begin{equation}
\label{eq:diagsmod3}
\begin{split}
	i\widetilde{\Gamma}_\lambda^{Z\mu\mu} &= \vcenter{\hbox{ \begin{tikzpicture}
    \begin{feynman}[scale=0.90,transform shape]
      \vertex (a) at (-0.3,0);
      \vertex (v1) at (1.5,0);
      \vertex (b) at (3.17,1.3);
      \vertex (bm) at (2.33,0.65);
      \vertex (c) at (3.17,-1.3);
      \vertex (cm) at (2.33,-0.65);
      \diagram* {
        (a) --[photon, edge label={$Z$}] (v1),
        (v1)--[fermion,edge label=$\chi$] (bm),
        (bm)--[fermion,edge label=$\mu^-$] (b),
        (v1)--[anti fermion,edge label'=$\chi$] (cm),
        (cm)--[anti fermion, edge label'=$\mu^+$] (c),
        (bm)--[scalar ,edge label=$S$] (cm)
        };
    \end{feynman}
\end{tikzpicture}}} + \vcenter{\hbox{\begin{tikzpicture}
    \begin{feynman}[scale=0.90,transform shape]
      \vertex (a) at (-0.3,0);
      \vertex (v1) at (1.5,0);
      \vertex (b) at (3.17,1.3);
      \vertex (bm) at (2.33,0.65);
      \vertex (c) at (3.17,-1.3);
      \vertex (cm) at (2.33,-0.65);
      \diagram* {
        (a) --[photon, edge label={$Z$}] (v1),
        (v1)--[fermion,edge label=$\chi$] (bm),
        (bm)--[fermion,edge label=$\mu^-$] (b),
        (v1)--[anti fermion,edge label'=$\chi$] (cm),
        (cm)--[anti fermion, edge label'=$\mu^+$] (c),
        (bm)--[scalar ,edge label=$A$] (cm)
        };
    \end{feynman}
\end{tikzpicture}}}+\vcenter{\hbox{\begin{tikzpicture}
    \begin{feynman}[scale=0.90,transform shape]
      \vertex (a) at (-0.3,0);
      \vertex (v1) at (1.5,0);
      \vertex (b) at (3.17,1.3);
      \vertex (bm) at (2.33,0.65);
      \vertex (c) at (3.17,-1.3);
      \vertex (cm) at (2.33,-0.65);
      \diagram* {
        (a) --[photon, edge label={$Z$}] (v1),
        (v1)--[scalar,edge label=$S$] (bm),
        (bm)--[fermion,edge label=$\mu^-$] (b),
        (v1)--[scalar,edge label'=$A$] (cm),
        (cm)--[anti fermion, edge label'=$\mu^+$] (c),
        (bm)--[fermion, edge label=$\chi$] (cm)
        };
    \end{feynman}
\end{tikzpicture}}}+\vcenter{\hbox{\begin{tikzpicture}
    \begin{feynman}[scale=0.90,transform shape]
      \vertex (a) at (-0.3,0);
      \vertex (v1) at (1.5,0);
      \vertex (b) at (3.17,1.3);
      \vertex (bm) at (2.33,0.65);
      \vertex (c) at (3.17,-1.3);
      \vertex (cm) at (2.33,-0.65);
      \diagram* {
        (a) --[photon, edge label={$Z$}] (v1),
        (v1)--[scalar,edge label=$A$] (bm),
        (bm)--[fermion,edge label=$\mu^-$] (b),
        (v1)--[scalar,edge label'=$S$] (cm),
        (cm)--[anti fermion, edge label'=$\mu^+$] (c),
        (bm)--[fermion, edge label=$\chi$] (cm)
        };
    \end{feynman}
\end{tikzpicture}}} ,\\
	i\widetilde{\Sigma}_\mu(\slashed{p}) &=\vcenter{\hbox{\begin{tikzpicture}
    \begin{feynman}[scale=0.90,transform shape]
      \vertex (a) at (0,0);
      \vertex (v1) at (1.43,0);
      \vertex (v2) at (2.87,0);
      \vertex (b) at (4.3,0);
      \diagram* {
        (a) --[fermion, edge label'=$\mu^-$] (v1),
        (v1) -- [scalar, half left, looseness=1.68, edge label=$S$] (v2),
        (v2)  -- [anti fermion, half left, looseness=1.68, edge label=$\chi$] (v1),
        (v2) --[fermion, edge label'=$\mu^-$] (b)
        };
    \end{feynman}
\end{tikzpicture}}}  +\vcenter{\hbox{\begin{tikzpicture}
    \begin{feynman}[scale=0.90,transform shape]
      \vertex (a) at (0,0);
      \vertex (v1) at (1.43,0);
      \vertex (v2) at (2.87,0);
      \vertex (b) at (4.3,0);
      \diagram* {
        (a) --[fermion, edge label'=$\mu^-$] (v1),
        (v1) -- [scalar, half left, looseness=1.68, edge label=$A$] (v2),
        (v2)  -- [anti fermion, half left, looseness=1.68, edge label=$\chi$] (v1),
        (v2) --[fermion, edge label'=$\mu^-$] (b)
        };
    \end{feynman}
\end{tikzpicture}}} .
\end{split}
\end{equation}
\begin{figure}[h!]
    \centering
    \includegraphics[scale=0.4]{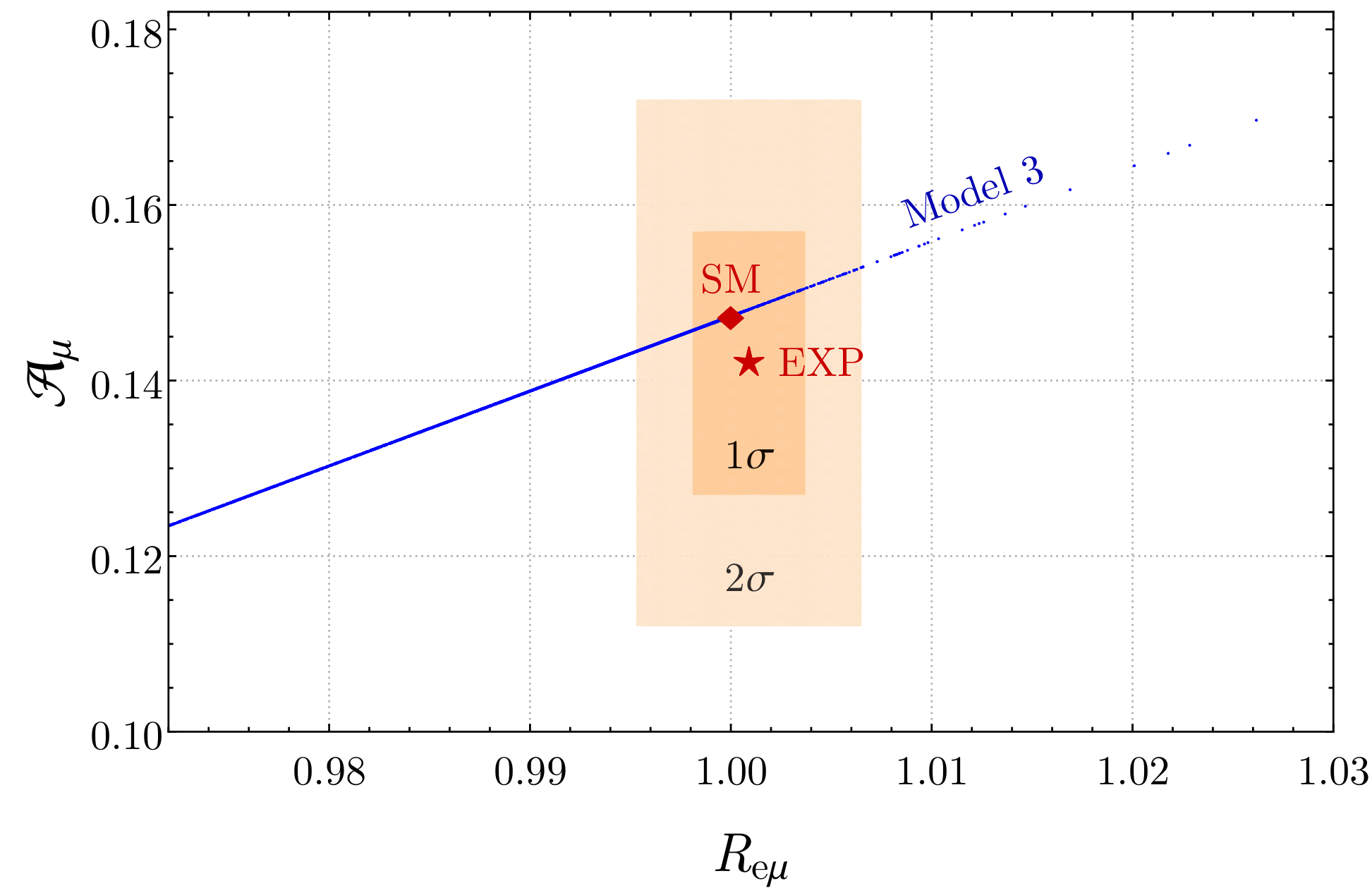}
    \caption{Representation of the $\mathcal{A}_\mu^0 - R_{e\mu}$ plane. The $1\sigma$ and $2\sigma$ experimental regions are shown in different shades of orange, together with a red star marking the experimental central value. The red diamond marks the SM prediction. The blue points show how Model 3 impacts the SM prediction.}
    \label{fig:mod3_Rem_Am}
\end{figure}

 In Fig. \ref{fig:mod3_Rem_Am}  we show in blue how the Model 3 points impact the SM prediction marked by the red diamond in the $\mathcal{A}_\mu^0 - R_{e\mu}$ plane. The $1\sigma$ and $2\sigma$ experimental regions are shown in different shades of orange, together with a red star marking the experimental central value. 
The relevant parameters to calculate the corrections to the couplings are the masses of the fermion, $m_\chi$, and of the scalars, $m_S$, $m_A$, and the coupling $y_\mu$. For Model 3, $m_S$ was varied from 5 to 100 GeV and $m_A$ from 100 to 1000 GeV. The limits for $m_\chi$ are the same as in Model 5. $\delta\widetilde{g}_R\propto m_\mu$ and is always several orders of magnitude smaller than $\delta \widetilde{g}_L$, which is the dominant term. By varying the parameters in the allowed ranges defined in~\cite{Capucha:2022kwo}, we can see that, unlike in Model 5, $\delta\widetilde{g}_L$ and $\delta\widetilde{g}_R$ can be both positive or negative. Once again it is possible to use $\mathcal{A}_\mu^0$ and $R_{e\mu}$ to constrain the parameter space.

\begin{figure}[h!]
    \centering
\hfill
\includegraphics[width=0.32\textwidth]{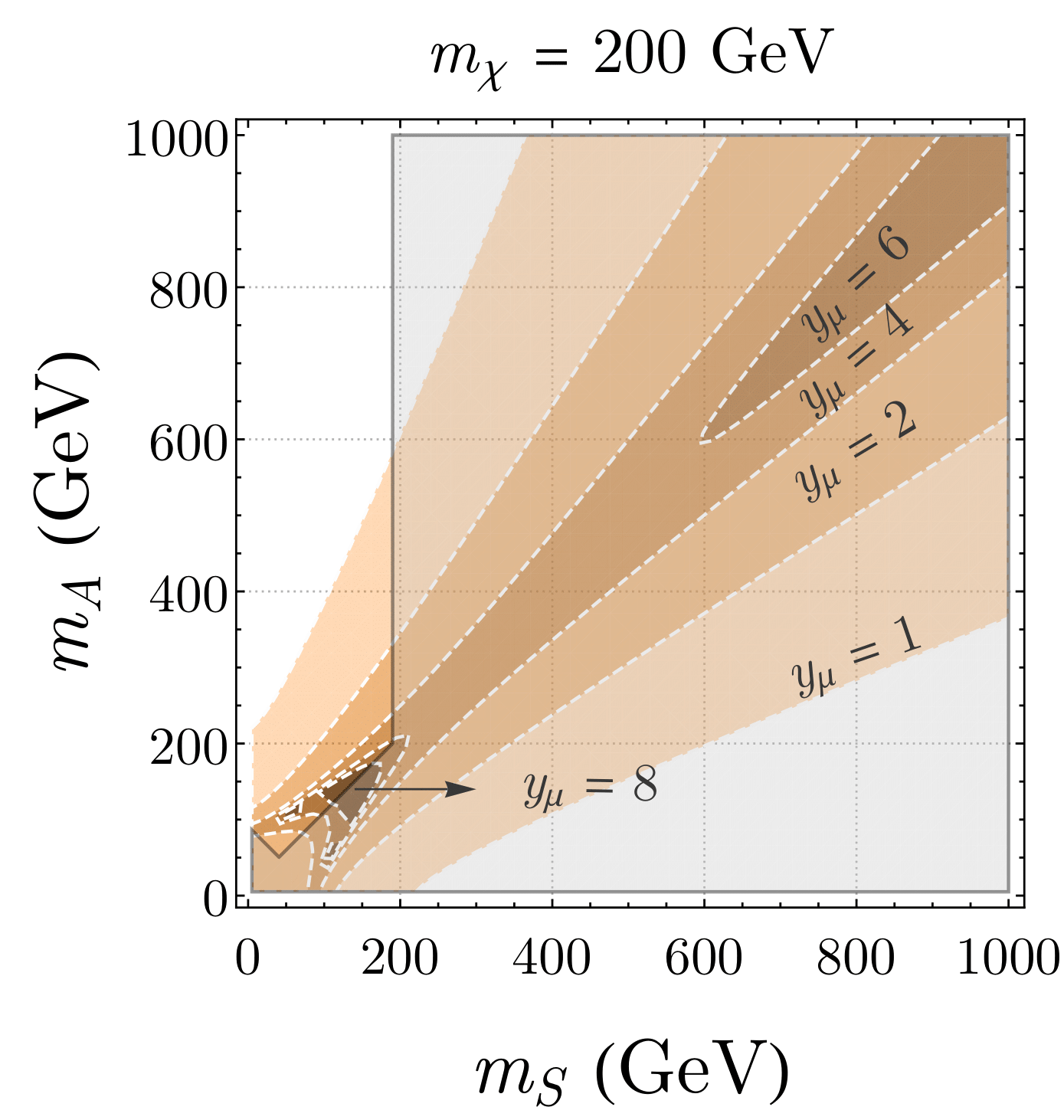}
\hfill
\includegraphics[width=0.32\textwidth]{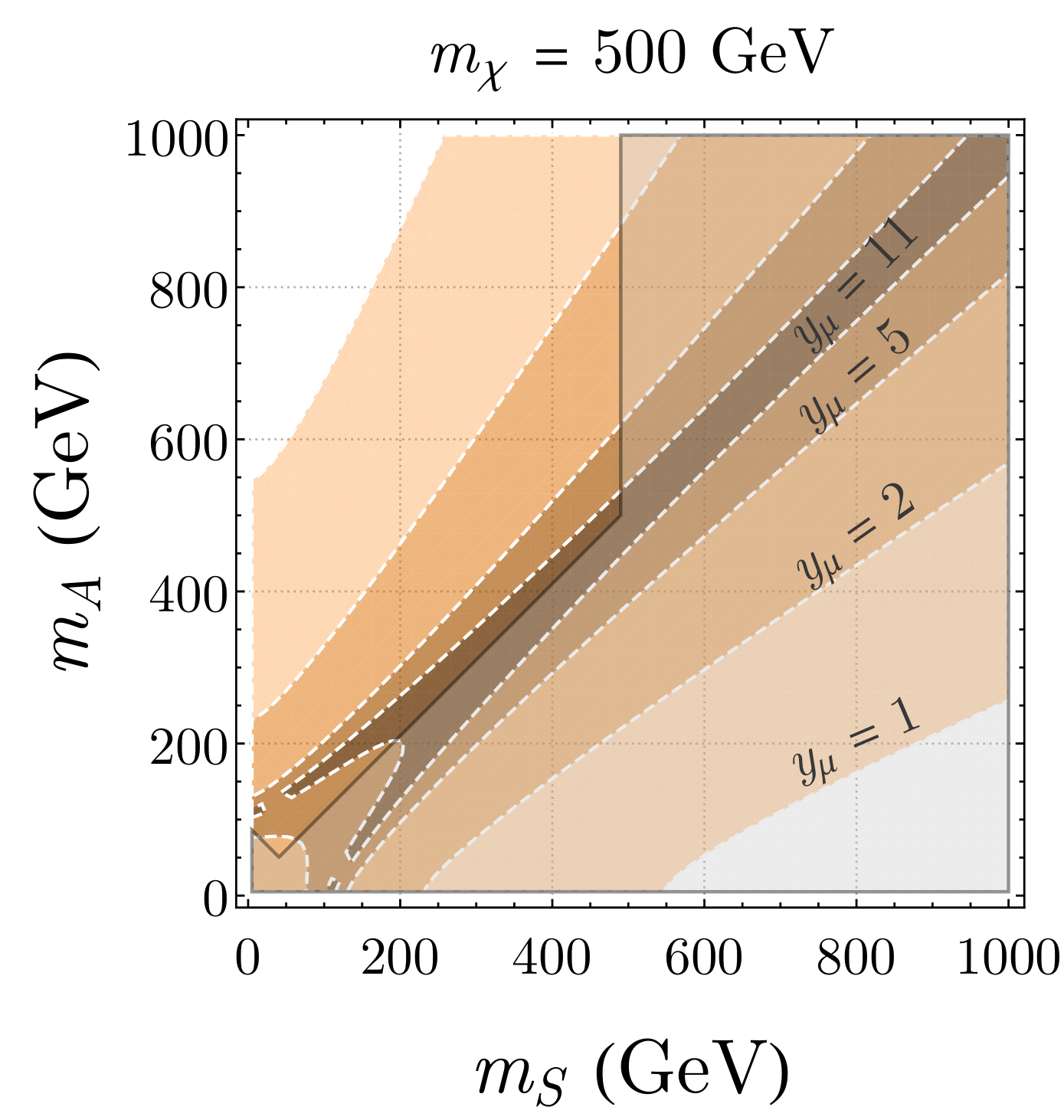}
\hfill
\includegraphics[width=0.32\textwidth]{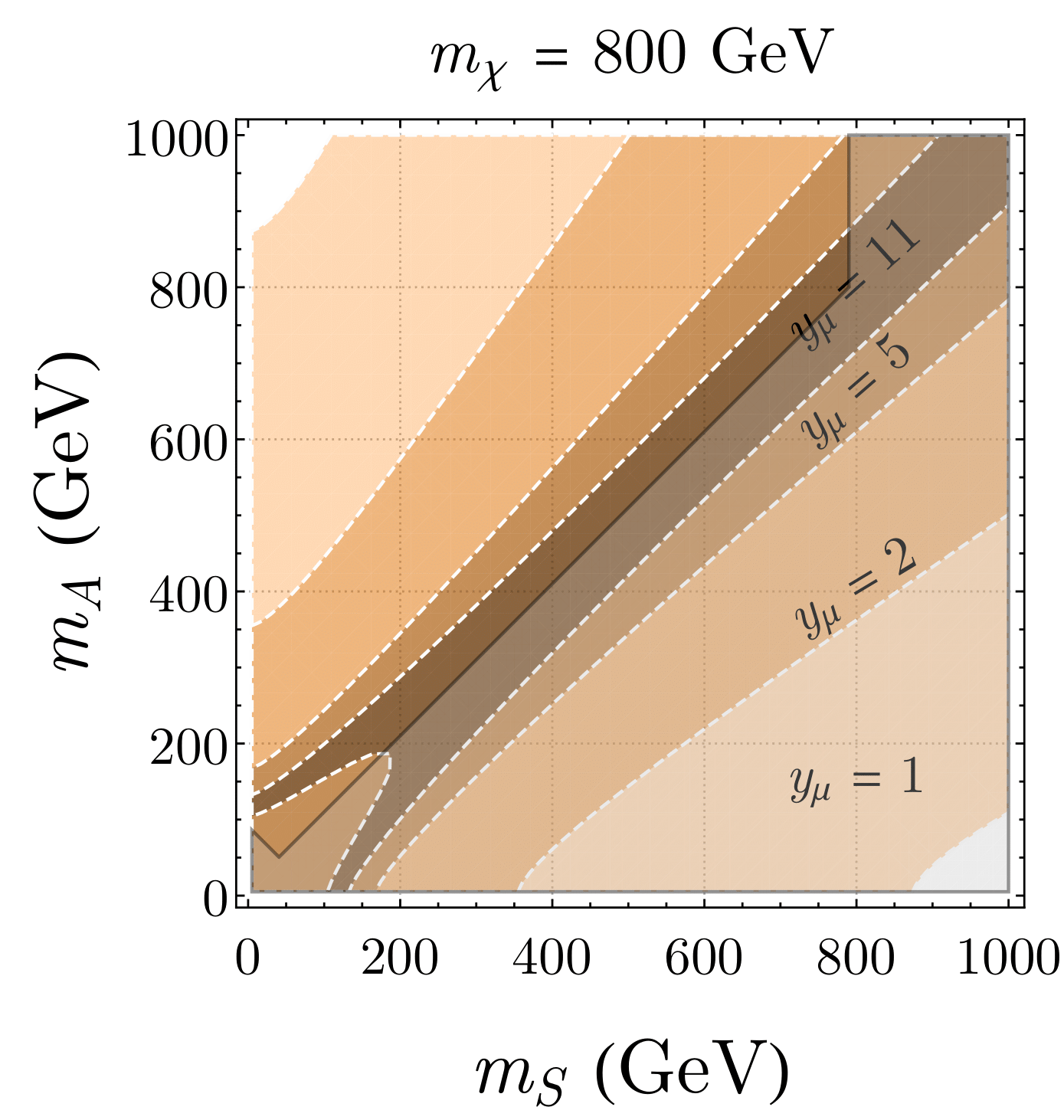}
\hfill
    \caption{$m_S-m_A$ projection plane with the values that keep $\mathcal{A}_\mu^0$ and $R_{e\mu}$ within the $1\sigma$ uncertainty ranges for $m_\chi=200,350$ and $700$ GeV, respectively. Darker shades of orange correspond to higher values for $y_\mu$ as indicated in the legend. Higher values of $y_\mu$ bring $m_S$ and $m_A$ closer together. Smaller values of $m_\chi$ decrease the allowed parameter space. The grey-shaded region is forbidden.}
    \label{fig:mod3_contour}
\end{figure}  

In Fig. \ref{fig:mod3_contour} we plot the 1$\sigma$ acceptance regions for $\mathcal{A}_\mu^0$ and $R_{e\mu}$ in the $m_S-m_A$ plane for different values of $y_\mu$ and $m_\chi$. The forbidden region satisfying $m_S > m_\chi + 10$ GeV, $m_S > m_A + 10$ GeV and $m_S + m_A < m_Z$ is gray-shaded. From these plots, we can conclude that in general the stronger the coupling, the closer the masses of the scalars must be. Also the smaller the values of $m_\chi$, the smaller the accepted region is. These are important results that show that on one hand the smaller gap between the mass of two scalars is preferred, 
especially for smaller values of $m_\chi$ and large values of the muon coupling. If further restrictions are imposed on even just one of the scalar masses limiting it to a small range, then it is expected that this would translate to a severe restriction of the allowed parameter space. 

\begin{figure}[h!]
     \centering
     \begin{subfigure}[b]{0.44\textwidth}
         \centering
         \includegraphics[width=\textwidth]{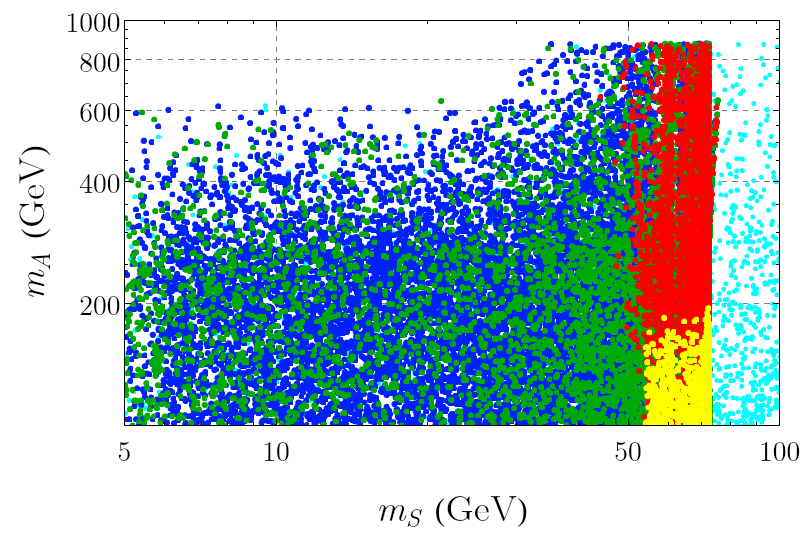}
         \caption{}
         \label{fig:mod3_a}
     \end{subfigure}
     \hfill
     \begin{subfigure}[b]{0.44\textwidth}
         \centering
         \includegraphics[width=\textwidth]{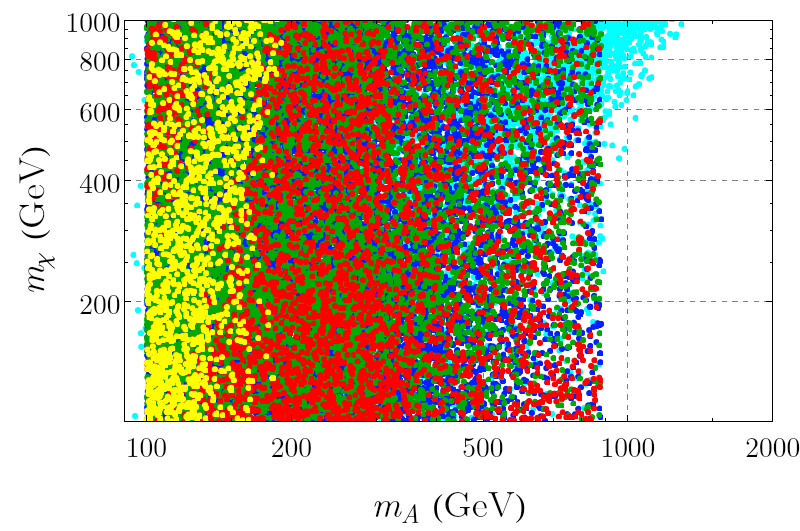}
         \caption{}
         \label{fig:mod3_b}
     \end{subfigure}
     \hfill
     \begin{subfigure}[b]{0.44\textwidth}
         \centering
         \includegraphics[width=\textwidth]{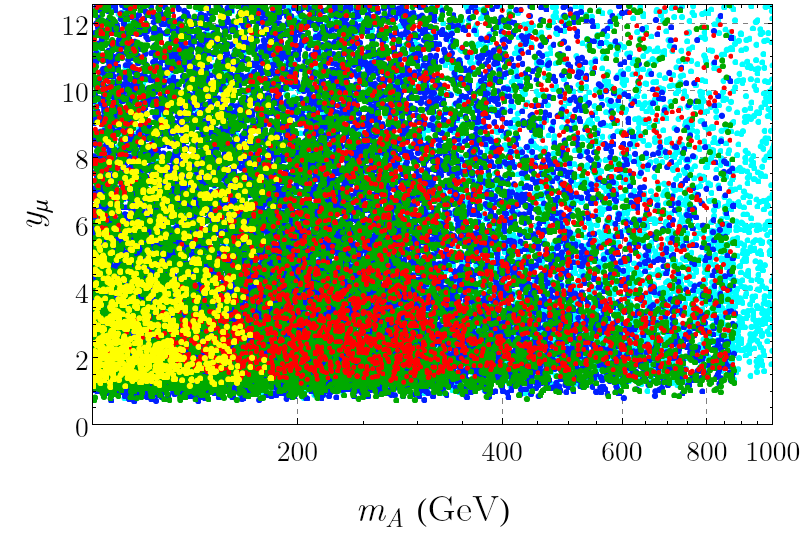}
         \caption{}
         \label{fig:mod3_c}
     \end{subfigure}
     \hfill
     \begin{subfigure}[b]{0.44\textwidth}
         \centering
         \includegraphics[width=\textwidth]{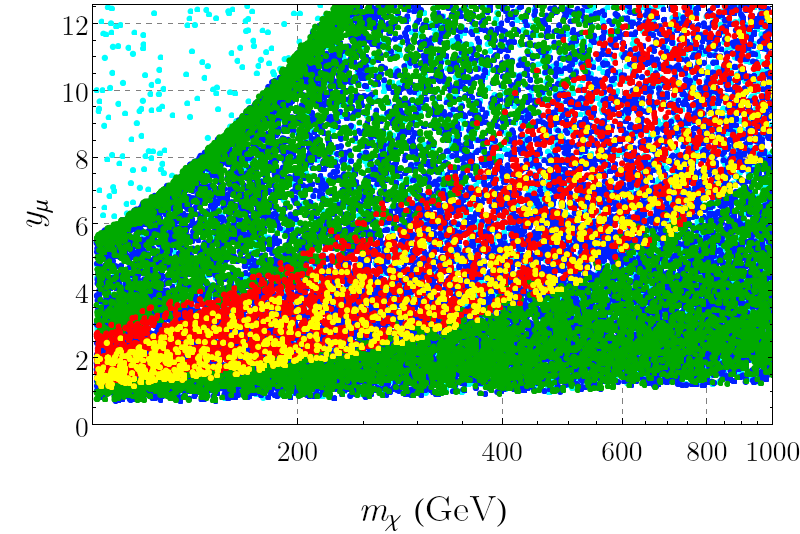}
         \caption{}
         \label{fig:mod3_d}
     \end{subfigure}     
        \caption{Allowed parameter space of Model 3 projected in the planes $m_S-m_A$ (top left), $m_A-m_\chi$ (top right), $m_A-y_\mu$ (bottom left) and $m_\chi-y_\mu$ (bottom right). The color scheme replicates the one used in \cite{PhysRevD.102.075009}. Only the yellow points agree with $\mathcal{A}_\mu^0$ and $R_{e\mu}$ at the $1\sigma$ level while satisfying all previous constraints. Every point agrees with $T$ to 2$\sigma$. Unlike in Model 5, the new muon-related restrictions can effectively limit the allowed parameter space.}
        \label{fig:mod3_all_plots}
\end{figure} 

\begin{figure}[h]
    \centering
    \includegraphics[width=0.44\textwidth]{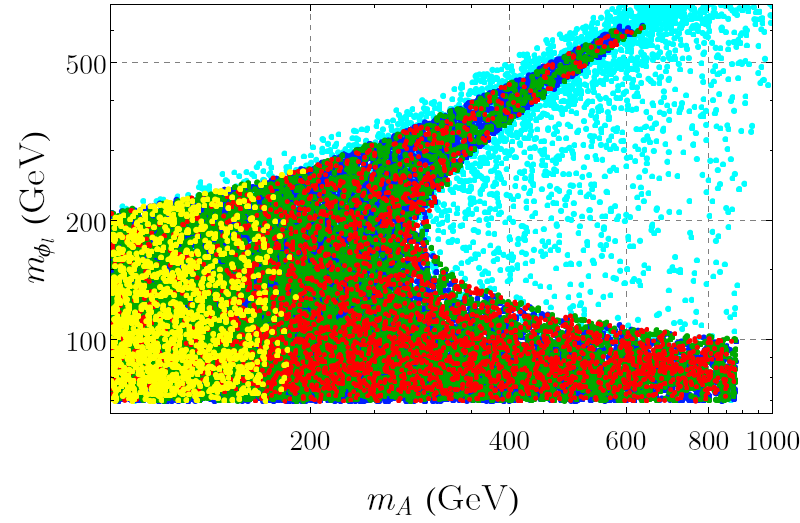}    
    \caption{Allowed parameter space of Model 3 projected in the plane $m_A-m_{\phi_l}$. The color scheme replicates the one used in \cite{PhysRevD.102.075009}. Only the yellow points agree with $\mathcal{A}_\mu^0$ and $R_{e\mu}$ at the $1\sigma$ level while satisfying all previous constraints. Every point agrees with $T$ to 2$\sigma$.}
    \label{fig:mod3_mA_mphil}
\end{figure} 

Finally, in Figs.\ref{fig:mod3_all_plots} and \ref{fig:mod3_mA_mphil} we review our results for Model 3 in view of the new constraints. We use the scan with the new results as already discussed for Model 5 and show four projections of the parameter space from the random scan. Once again the yellow points correspond to the points that verify all dark matter and flavor-related constraints and the new muon-related constraints to 1$\sigma$. Every point shown in the plots is in accordance with the experimental result for the $T$ parameter to 2$\sigma$. This restriction has a visible consequence in the $m_S-m_A$ plane, where points where the difference between $m_A$ and $m_S$ is too large to obey this condition are excluded. In stark contrast to what we got for Model 5, in Model 3 $\mathcal{A}_\mu^0$ and $R_{e\mu}$ can significantly limit the allowed regions.  As before, $R_{e\mu}$ provides the strongest constraint, with all points accepted by the condition defined with $\mathcal{A}_\mu$ to 1 $\sigma$. As noted in \cite{Capucha:2022kwo}, in Model 3 the dark matter constraints limit $m_S$ to be bigger than 42 GeV and smaller than 76 GeV. This combined with the muon-related constraints means the allowed values for $m_A$ are much more restricted than indicated by the $g-2$ constraints. The dark matter constraints end up indirectly limiting the allowed values of $m_A$. This can be seen in the $m_S-m_A$, $m_A-m_\chi$ and $m_A-y_\mu$ projections, where a large portion of the red points are excluded by $R_{e\mu}$. The yellow points are concentrated as expected in the smaller mass regions. In the plane $m_A-y_\mu$ we can see that as $y_\mu$ increases, the allowed range of $m_A$ gets closer to the allowed interval of values for $m_S$.  This is consistent with the behavior observed in Fig. \ref{fig:mod3_contour}. As in Model 5, the muon-related constraints now reject a small strip of previously accepted space in the $m_\chi-y_\mu$ plane. These observables also provide a much stronger constraint to the mass of $\phi_l$ than the $g-2$ results, with $m_{\phi_l}$ now allowed to be at most 265 GeV. We can summarize the new bounds on the parameters of Model 3 as
\begin{align*}
&&		m_A &\in [100,194]	\textup{ GeV},			&		y_\mu &\in [1,4\pi] , 			&&\\
&&		m_\chi &> 101.2 \textup{ GeV},		&		m_S &\in [50,75] \textup{ GeV}. &&\\
&& m_{\phi_l} &\in [70,265]	\textup{ GeV}, & &&
\end{align*}

\begin{figure}[h!]
     \centering
         \centering
         \includegraphics[width=0.38 \textwidth]{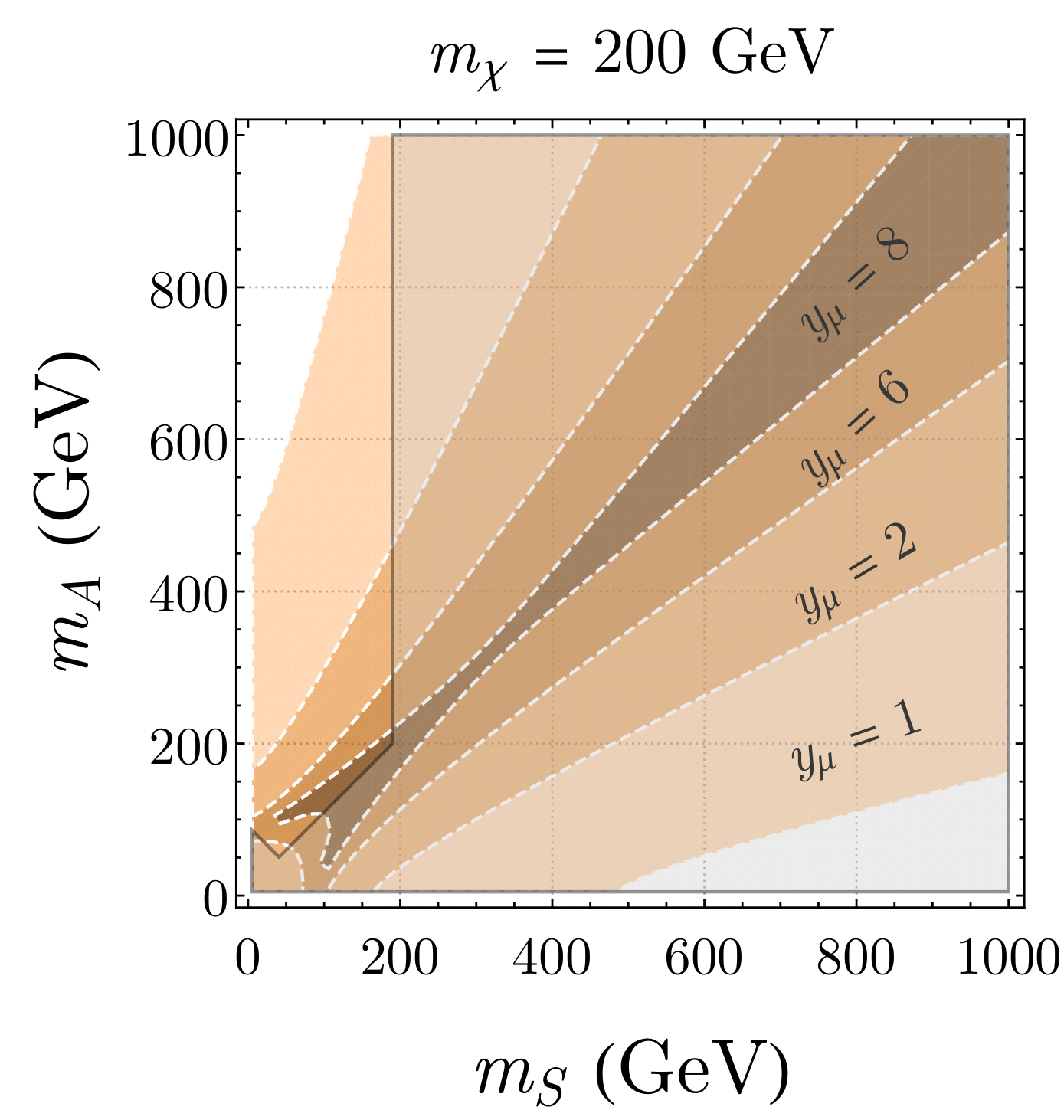}
         \hspace{1cm}
         \includegraphics[width=0.38 \textwidth]{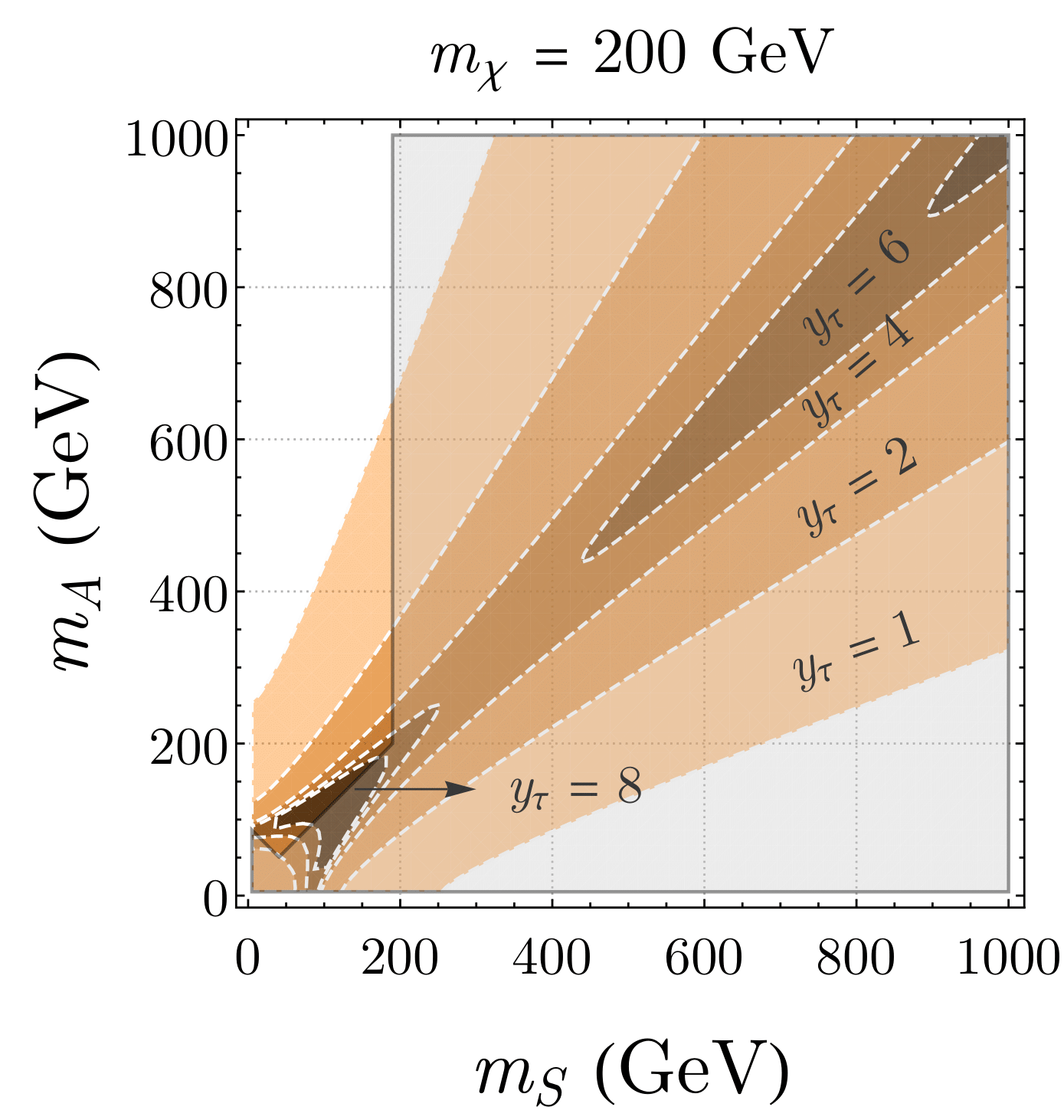}
        \caption{$m_S-m_A$ projection plane with the values that keep $\mathcal{A}_\mu^0$ and $R_{e\mu}$ (left panel), and $\mathcal{A}_\tau^0$ and $R_{e\tau}$ (right panel)  within the $2 \sigma$  uncertainty ranges for $m_\chi=200$ GeV. Darker shades of orange correspond to higher values for $y_\mu$ 
        and $y_\tau$ as indicated in the legend. The grey-shaded region is forbidden.}
        \label{fig:mmtt}
\end{figure}  

We end this section with a comment on the use of the process $Z \to \tau^+ \tau^-$.  In Fig.~\ref{fig:mmtt}  we present the $m_S-m_A$ projection plane with the regions that keep $\mathcal{A}_\mu^0$, and $R_{e\mu}$ (left panel), and $\mathcal{A}_\tau^0$ and $R_{e\tau}$ (right panel) within the $2 \sigma$ uncertainty ranges, for $m_\chi=200$ GeV. Again the darker shades of orange correspond to higher values for $y_\mu$ and $y_\tau$ as shown in the legend. Both calculations are the same from the theoretical side and there is no noticeable difference coming from the different masses. However, a direct comparison between the two plots should be done with care, since, unlike the muon case, the SM prediction for $R_{e \tau}$ is outside the $1\sigma$ experimental uncertainty range. Therefore, a correlation between constraints on the tau and experimental precision cannot be made.

\section{Conclusions \label{sec:conclusions}}

We have proposed a set of precision muon-related observables that serve as a tool to constrain new physics models. Taking advantage of the precision measurements performed by the LEP collaborations on the $Z$-boson pole, it is possible to perceive the contribution of new physics to measured SM processes via quantum corrections. The asymmetry parameter ($\mathcal{A}_\mu^0$) and the ratio of lepton hadronic decay rates ($R_{e\mu}$) are extracted from the $e^+ e^- \to \mu^+ \mu^-$ process. With these observables at hand, we can find the limits on the variations of the $Z \mu^+ \mu^-$ couplings when one-loop contributions from new physics are included. We provide a simple recipe that can be used for any extension of the SM.

We have discussed in detail the impact of the new bounds on two particular models in which a new Dark Sector is added to the SM. The new sector includes a vector like fermion and two scalars. In one scenario, dubbed Model 5 the scalars are $SU(2)$ singlets and the fermion is an $SU(2)$ doublet. In the other scenario, Model 3, the scalars are $SU(2)$ doublets and the fermion is an $SU(2)$ singlet. While in the first scenario, the new constraints are very mild, for Model 3 they constitute a real bound on the low mass region of the scalars. This is especially important when combined with other constraints

These bounds are particularly relevant for models with an enhanced coupling to the muons as is the case of the ones that attempt to solve the $g-2$ anomaly via new physics loop contributions.

\vspace*{1cm}
\subsubsection*{Acknowledgments}
\noindent
GL and RS are partially supported by the Portuguese Foundation for Science and Technology (FCT) under Contracts no. UIDB/00618/2020, UIDP/00618/2020, CERN/FIS-PAR/0025/2021 and CERN/FIS-PAR/0021/2021.
The work of A.M. and J.P.S. is supported in part by FCT under Contracts CERN/FIS-PAR/0008/2019, PTDC/FIS-PAR/29436/2017, UIDB/00777/2020, and UIDP/00777/2020; these projects are partially funded through POCTI (FEDER),
COMPETE, QREN, and the EU.

\vspace*{1cm}
\bibliographystyle{h-physrev}
\bibliography{zmumu.bib}

\end{document}